\documentclass[oldversion]{aa}
\usepackage{natbib}
\usepackage{graphicx}
\usepackage{amssymb}
\usepackage{epsfig}
\usepackage{aasmacros}
\usepackage{hyperref}
\begin{document}
\def\sizex{16.0 cm}
\def\bigx{10.0 cm}
\def\smallerxsize{7.0 cm}
\def\smallxsize{10.0 cm}
\def\smallysize{12.0 cm}
\title {Clustering properties of a type-selected volume-limited\\ sample of
  galaxies in the CFHTLS\thanks{Based on
    observations obtained with MegaPrime/MegaCam, a joint project of
    CFHT and CEA/DAPNIA, at the Canada-France-Hawaii Telescope (CFHT)
    which is operated by the National Research Council (NRC) of Canada,
    the Institut National des Science de l'Univers of the Centre
    National de la Recherche Scientifique (CNRS) of France, and the
    University of Hawaii. This work is based in part on data products
    produced at TERAPIX and the Canadian Astronomy Data Centre as part
    of the Canada-France-Hawaii Telescope Legacy Survey, a
    collaborative project of NRC and CNRS.}}
\offprints {H.\ J.\ McCracken} 
\date { }
\titlerunning{Galaxy clustering in the CFHTLS}
\authorrunning{McCracken et al.}

\abstract{We present an investigation of the clustering of the faint
  ($i'_{AB}<24.5$) field galaxy population in the redshift range
  $0.2<z<1.2$. Using 100,000 precise photometric redshifts extracted
  from galaxies in the four ultra-deep fields of the Canada-France
  Legacy Survey, we construct a set of volume-limited galaxy
  samples. We use these catalogues to study in detail the dependence
  of the amplitude $A_w$ and slope $\delta$ of the galaxy correlation
  function $w$ on absolute $M_B$ rest-frame luminosity, redshift, and
  best-fitting spectral type (or, equivalently, rest-frame
  colour). Our derived comoving correlation lengths for
  magnitude-limited samples are in excellent agreement with
  measurements made in spectroscopic surveys. Our principal
  conclusions are as follows: 1. The comoving correlation length for
  all galaxies with $-19<M_B-5\log h<-22$ declines steadily from
  $z\sim0.3$ to $z\sim1$.  2. At all redshifts and luminosity ranges,
  galaxies with redder rest-frame colours have clustering amplitudes
  between two and three times higher than bluer ones. 3. For both the
  red and blue galaxy populations, the clustering amplitude is
  invariant with redshift for bright galaxies ($-19<M_B-5\log
  h<-22$). 4. At $z\sim0.5$ for less luminous galaxies with $M_B-5\log
  h\sim-19$ we find higher clustering amplitudes of $\sim
  6h^{-1}$~Mpc. 5. The relative bias between redder and bluer
  rest-frame populations increases gradually towards fainter
  magnitudes. Among the principal implications of these results is
  that although the full galaxy population traces the underlying dark
  matter distribution quite well (and is therefore quite weakly
  biased), redder, older galaxies have clustering lengths which are
  almost invariant with redshift, and by $z\sim1$ are quite strongly
  biased.}
\author{H.\ J.\ McCracken \inst{1} \and O. Ilbert \inst{2} \and Y.
  Mellier \inst{1} \and
  E. Bertin \inst{1} \and L.  Guzzo\inst{3}\and \\
  S. Arnouts\inst{4,5}, O. Le F\`evre \inst{4} \and G. Zamorani
  \inst{6}} \institute{Institut d'Astrophysique de Paris, UMR7095
  CNRS, Universit\`e Pierre et Marie Curie, 98 bis Boulevard Arago,
  75014 Paris, France \and Institute for Astronomy, 2680 Woodlawn
  Dr.,University of Hawaii, Honolulu, Hawaii, 96822\and
  INAF-Osservatorio Astronomico di Brera, via Bianchi 46, I-23807
  Merate (LC), Italy\and Laboratoire d'Astrophysique de Marseille, BP
  8, Traverse du Siphon, 13376 Marseille Cedex 12, France\and
  Canada-France-Hawaii telescope, 65-1238 Mamalahoa Highway, Kamuela,
  HI 96743 \and Osservatorio Astronomico di Bologna, Via Ranzani 1,
  40127 Bologna , Italy}

\maketitle
\section{Introduction}
In the cold dark matter model structures grow hierarchically under the
influence of gravity. Galaxies form inside ``haloes'' of dark matter
\citep{1978MNRAS.183..341W}. Because these haloes can only form at the
densest regions of the dark matter distribution, the distribution of
galaxies and dark matter is not the same; the more strongly clustered
galaxies are said to be ``biased''
\citep{1984ApJ...284L...9K,1986ApJ...304...15B} with respect to the
dark matter distribution. This relationship between dark and luminous
matter provides important information concerning the galaxy formation
process and tracing the evolution of bias as a function of scale and
mass of the hosting dark matter halo is one of the key objectives of
observational cosmology. On large scales, ($>10 h^{-1}$~Mpc) structure
growth is largely driven by gravitation (where we measure the
correlations between separate haloes of dark matter); however on
smaller scales ($<1~h^{-1}$) non-linear effects generally associated
with galaxy formation dominate the structure formation process. In
this paper we must bear in mind that although we measure a clustering
signal to around $0.1\deg$ at $z\sim1$ this corresponds to around
$\sim 3h^{-1}$~Mpc and $\sim 2h^{-1}$~Mpc at $z\sim0.5$, which means
that our observations are mostly in non-linear to strongly non-linear
regimes where environmental effects play an important role in the
evolution of structure.

On linear scales, as theory and simulations have shown (for example
\cite{1998ApJ...499...20J} or \cite{2004ApJ...601....1W}), the
clustering amplitude of dark matter decreases steadily with
redshift. If galaxies perfectly traced the dark matter component, then
their clustering amplitudes would decrease at each redshift slice, in
step with the underlying dark matter. However, as the galaxy
distribution is biased, stellar evolution intervenes to complicate
this picture; in effect, the actual measured clustering amplitudes are
a complicated interplay between the underlying dark matter component
and how well the luminous matter traces this galaxy distribution, or
how efficiently galaxies form. Understanding fully the evolution of
galaxy clustering requires, therefore, some insights into the galaxy
formation process. Thanks to large spectroscopic redshift surveys we
now have a much more complete picture of the evolution of the galaxy
luminosity function with redshift \citep{2005A&A...439..863I} and how
the fraction of galaxy types evolves with redshift
\citep{2006A&A...455..879Z}. For example, \cite{2005A&A...439..863I},
using first-epoch data from the VVDS redshift survey have shown that
the luminosity function brightens considerably between $z=0.3$ and
$z=1$, with $M*$ increasing by one or two magnitudes at $z\sim1$. We
must take this into account when comparing clustering amplitudes
measured at the same absolute luminosities in different redshift
ranges.

In the local Universe, million-galaxy redshift surveys have greatly
expanded our knowledge of galaxy clustering at low redshift. We now
have a broad idea how the distribution of galaxies depends on their
intrinsic luminosity and spectral type
\citep{2001MNRAS.328...64N,2002MNRAS.332..827N,2005ApJ...630....1Z}.
In general, these works have shown to a high precision that at the
current epoch more luminous galaxies are more clustered than faint
ones, and that similarly redder objects have higher clustering
amplitudes than bluer ones. Other works have shown that slope of the
galaxy correlation depends on spectral type
\citep{2003MNRAS.344..847M}. These studies have indicated that, in
general, more luminous, redder, objects are more strongly clustered
than bluer, fainter galaxies. Some studies have also related physical
galaxy properties, such a total mass in stars, with the clustering
properties \citep{2006MNRAS.368...21L}. But how do these relationships
change with look back time?

At intermediate redshifts ($z\sim1$), however, our knowledge is still
incomplete. Multi-object spectrographs mounted on ten-metre class
telescopes have made it possible to construct samples of a few
thousand galaxies.  The first studies investigating galaxy clustering
as a function of the object's rest frame luminosity and colour for
large galaxy samples at $z\sim1$ have now appeared
\citep{2006A&A...452..387M,2006A&A...451..409P,2006ApJ...644..671C}.
Unfortunately, these surveys typically contain $\sim10^3$ galaxies,
which are enough to select objects either by type and absolute
luminosity, but not, for instance, to apply both cuts simultaneously.

These works confirm some of the broad trends seen at lower redshift
and with magnitude-limited samples \citep{2005A&A...439..877L} but are
still not quite large enough to investigate in detail how galaxy
clustering depends simultaneously on more than one galaxy property.
For example, one may investigate the dependence of clustering
amplitude within a volume-limited sample \citep{2006A&A...452..387M},
but one may not, as yet, investigate simultaneously samples selected
by type and absolute luminosity. Unfortunately, even with efficient
wide-field multi-object spectrographs, gathering redshift samples of
thousands of galaxies at redshift of one or so requires a significant
investment of telescope time.

Photometric redshifts offer an exit from this impasse, and represent a
middle ground between simple studies using imaging data with magnitude
or colour-selected samples and spectroscopic surveys. Several attempts
have been made in the past to carry out galaxy clustering studies with
photometric redshifts, mostly using the Hubble deep field data sets
\citep{2002MNRAS.329..355A,2001ApJ...548..127T,1999MNRAS.306..988M}.
However, such works either suffered from sampling and cosmic variance
issues or poorly-controlled systematic errors. The advent of
wide-field mosaic cameras like MegaCam \citep{2000SPIE.4008..657B} has
made it feasible for the first time to construct samples of tens to
hundreds of thousands of galaxies from $z\sim0.2$ all the way to
$z\sim1$ and beyond. Two key advances have made this possible;
firstly, rigorous quality control of photometric data, and secondly,
the availability of much larger, reliable training samples reaching to
faint ($i'\sim24$) magnitudes

In this paper we will describe measurements of galaxy clustering
derived from a large sample of galaxies with accurate photometric
redshifts in Canada-France legacy survey (CFHTLS) deep fields. These
fields have been observed repeatedly since the start of survey
operations in June 2003 as part of the on-going SNLS project
\citep{2006A&A...447...31A} and consequently each filter has very long
integration times (for $r$ and $i$ bands the total integration time in
certain fields is over 100 hours).  A full description of our
photometric redshift catalogue can be found in
\cite{2006A&A...457..841I}.  Containing almost 100,000 galaxies to
$i'<24.5$ we are able to divide our sample by redshift, absolute
luminosity and spectral type.  These photometric redshifts have been
calibrated using 8,000 spectra from the VIRMOS-VLT deep survey (VVDS;
\citet{2005A&A...439..845L}). In addition, our sample has sufficient
volume to provide reliable measurements of galaxy clustering
amplitudes at redshifts as low as $z\sim0.2$; and we are thus able to
follow the evolution of galaxy correlation lengths over a wide
redshift interval.  In the lower redshift bins, the extremely deep
CFHTLS photometry means it is possible to measure clustering
properties of a complete sample of objects as faint as $M_B-5
\log(h)\sim-18$ (at $z\sim0.2$ we have large numbers of very faint
objects with $M_B-5\log(h)\sim-15$ although we do not consider them
here).  Moreover, by using all four independent deep fields of the
Canada-France legacy survey we are able to robustly estimate the
amplitude of cosmic variance for each of our samples.

Our objective in this paper is to determine, first of all, how the
observed properties of galaxies determines their clustering.  We are
able to carry out such an investigation of galaxy clustering strength
for samples of galaxies selected independently in absolute luminosity,
rest-frame colour and redshift.

Our paper is organised as follows: in
Section~\ref{sec:prep-catal-comp} we describe how our catalogues were
prepared and how we computed our photometric redshifts; in
Section~\ref{sec:meas-galaxy-clust} we describe how we measure galaxy
clustering in our data; our results are presented in
Section~\ref{sec:results}. Finally, our discussions and conclusions
are presented in Section~\ref{sec:disc-results-comp}.  In this work we
divide the CFHTLS galaxy samples in three principal ways: first of
all, we consider simple magnitude limited samples, divided by bins of
redshift (described in Section~\ref{sec:magn-limit-sampl}); next, at
two fixed redshift ranges, we consider galaxy samples selected by
absolute luminosity and type (Section \ref{sec:lumin-deped-clust});
and lastly, at a range of redshift bins and for the same slice in
absolute luminosity, we consider galaxies selected by type (presented
in Section~\ref{sec:volume-type-depend}).

Throughout the paper, we use a flat lambda cosmology ($\Omega_m~=~0.3$,
$\Omega_\Lambda~=~0.7$) and we define
$h~=~H_{\rm0}/100$~km~s$^{-1}$~Mpc$^{-1}$. Magnitudes are given in the
AB system unless otherwise noted.

\section{Catalogue preparation and photometric redshift computation}
\label{sec:prep-catal-comp}

We now describe the preparation of the photometric catalogues used to
derive our photometric redshifts. Although our input catalogue has
already been released to the community as part of the CFHTLS-T0003
release (hereafter ``T03''), no extensive description of the catalogue
processing has yet appeared in the literature; for completeness we
provide a brief outline of the principal processing steps in this
Section.

These photometric catalogues were released by the TERAPIX data centre
to the Canadian and French communities as part of the T03 release and
have been made public world-wide one year later. They comprise
observations taken with the MegaCam wide-field mosaic camera
\citep{2000SPIE.4008..657B} at the Canada-France-Hawaii telescope
between June 1st, 2003 and September 12th, 2005. Full details of these
observations, data reductions, catalogue preparation and quality
assessment steps can be found on the TERAPIX web
pages\footnote{\url{http://terapix.iap.fr/rubrique.php?id_rubrique=208}},
however, we now outline the principal steps in data reductions and
catalogue preparation.

\subsection{Production of stacked images}
\label{sec:prod-stack-imag}

MegaCam is a wide-field CCD mosaic camera consisting of 36 thinned EEV
detectors mounted at the prime focus of the 3.6m Canada France Hawaii
Telescope on Mauna Kea, Hawaii. The detectors are arranged in two
banks.  The nominal pixel scale at the centre of the detector is
0.186$\arcsec$/pixel; the size of each detector pixel is $13.5\mu$. All
observations for the CFHTLS are taken in queue-scheduling mode.  Each
of the four fields presented in this paper have been observed in all
five MegaCam broad-band filters primarily for the supernovae legacy
survey. After pre-processing (bias-subtraction and flat-fielding) at
the CFHT, images are transferred to the Canadian astronomy data centre
(CADC) for archiving, and thence to TERAPIX at the IAP in Paris for
processing. At TERAPIX, the data quality assessment tool ``QualityFITS''
is run on each image, which provides a 'report card' in the form of a
HTML page containing information on galaxy counts, stellar counts, and
the point-spread function for each individual image.  Catalogues and
weight-maps are also generated.  At this point each image is also
visually inspected and classified.  

In the classification process galaxies are divided into four grades
according to seeing and associated image features (for instance, if
the telescope lost tracking or other artifacts were present). Only the
two highest-quality grades are kept for subsequent analysis.

After all images have been inspected, and bad images rejected, an
astrometric and photometric solution is computed using the TERAPIX
tool \texttt{scamp} which computes a solution simultaneously for all
filters \citep{2006ASPC..351..112B}. Finally, this astrometric
solution is used to re-project and co-add all images (and weights) to
produce final stacked image. All of these steps are managed from an
web-based pipeline environment. The internal r.m.s. astrometric
accuracy over the entire MegaCam field of view is always less than one
MegaCAM pixel ($0.186\arcsec$)

\subsection{Quality assessment}
\label{sec:quality-assessment}

Galaxy number counts, stellar colour-colour plots and incompleteness
measurements have been calculated for all four deep stacks in all five
bands. By examining the position of the stellar locus in each field in
the $u-g$ vs $g-r$ and $g-r$ vs $r-i$ colour-colour planes we see that
the photometric zero-point accuracy field to-field is $\sim 0.03$ or
better. Detailed comparisons between CFHTLS-wide survey fields and
overlapping Sloan Digital Sky Survey fields show systematic errors of
a comparable amplitude. This degree of photometric precision is
essential if we are to compute accurate photometric redshifts. A full
list of the characteristics of release T03 can be found at {\tt \url
  {http://terapix.iap.fr/cplt/tab\_t03ym.html}}

\subsection{Catalogue generation}
\label{sec:catalogue-generation}

Once images have been resampled and median-combined for each field we
use \texttt{swarp} to produce a ``chi-squared'' detection image
\citep{1999AJ....117...68S} based on the $g'$, $r'$ and $i'$ stacks (the
pixel scale on each image in all fields and colours is fixed to
$0.186\arcsec$/pixel.  Next, \texttt{sextractor}
\citep{1996A&AS..117..393B} is executed in ``dual-image'' mode on all
stacks using the chi-squared image as the detection image. This method
``automatically'' produces matched catalogues between each stack as in
all cases the detection image remains the same. We note that, given
the strict criterion on the image seeing used to select input images
in the CFHTLS stacks, all deep stacks are approximately
seeing-matched, with the median seeing on each final stack in each
band of $\sim1\arcsec$. This means one can safely use dual-mode detection.
We use Sextractor's \texttt{mag\_auto} Kron-like ``total'' galaxy
magnitudes \citep{1980ApJS...43..305K}. At faint magnitudes, where the
error on the Kron radius can be large, our total magnitudes revert to
simple $3\arcsec$ diameter aperture magnitudes. After the extraction
of catalogues redundant information is removed from each band and a
``flag'' column is added to the catalogues containing information
about the object compactness using the ``local'' measurement of the
object's half-light radius \citep{2003A&A...410...17M}. 
A mask file, generated automatically and fine-tuned by hand, is
used to indicate areas near bright stars or with lower cosmetic
quality, and this information is incorporated in the object
flag. Objects used in the subsequent scientific analysis are those
which do not lie in these masked regions, are not saturated, and are
not stars.

\subsection{Photometric redshift computation and accuracy}
\label{sec:comp-phot-redsh}

A full description of our method used to compute photometric redshifts
is given in \cite{2006A&A...457..841I}. Briefly we use a two-step
optimisation process based on firstly the bright sample (to set the
zero-points) and the full sample (which optimises the templates). This
new template set is then used to compute photometric redshifts in all
four fields. In this paper we consider photometric redshifts computed
using only the five CFHTLS filters ($u^*griz$). This is true even in
fields where additional photometric information is available (for
example, CFHTLS-D1 field where there is supplementary $BVRIJK$
photometry). This approach was taken to ensure that field-to-field
variation in photometric redshift accuracy as a function of redshift
was kept to a minimum.  Our photometric redshifts are essentially
identical to those presented in \citeauthor{2006A&A...457..841I} with
the exception that in the D2 field we use additional ultra-deep $u*$
imaging kindly supplied by the COSMOS consortium; this serves to
equalise the $u*$ integration time between the fields. We separate
stellar sources from galaxies by using a combination of sextractor
\texttt{flux\_radius} parameter and the best fitting spectral
template.

We emphasise that a key aspect of our photometric redshifts is that
extensive comparisons have been made with large database of
spectroscopic redshifts \citep{2005A&A...439..877L}. In particular, we
draw attention to Figures 9 and 10 of \citeauthor{2006A&A...457..841I}
which show photometric redshift accuracy and the fraction of
catastrophic errors as a function of redshift. For galaxies with
$i'<24$ in the redshift range $0.2<z<1.2$ the photometric redshift
accuracy in the D1 field, expressed as $\sigma_{\Delta z}/(1+z)$, is
always less than 0.06; in the redshift range $0.2<z<0.6$ it is less
than 0.04.  Catastrophic errors are defined as the number of galaxies
with $|z_s-z_p|/(1+z_s)>0.15$ where $z_s$ is the spectroscopic
redshift and $z_p$ the photometric redshifts. From Figure 10 in
\citeauthor{2006A&A...457..841I} we can see that the fraction $\eta$
of objects with catastrophic redshift errors is better than $5\%$ in
the redshift range $0.2<z<1.2$ for objects with $22.5<i_{AB}<24.0$.

Although there are smaller numbers of spectroscopic redshifts in the
other fields, some useful comparisons can be made; using 364
publicly-available spectra from the DEEP1 project, Figure 14 in
\citeauthor{2006A&A...457..841I} shows that the dispersion $\delta
z/(1+z_s)$ is $0.03$ in the redshift interval $0.2<z<1.2$. In the
D2 field we have carried out an additional comparison with
spectroscopic redshifts obtained by J. P. Kneib and collaborators in
the context of the COSMOS project. This test, making use of 335 $i'<24$
spectroscopic redshifts, shows that, once again, in the interval
$0.2<z<1.2$, our photometric redshift errors $\delta z/(1+z_s)$ are
$\sim0.035$.

During the preparation of this article, an independent comparison has
been carried out by members of the DEEP2 team between their large
spectroscopic sample and the CFHTLS-T03 photometric redshifts
presented here. They find an excellent agreement between, comparable
to the values presented here for the other fields, for more than
20,000 galaxies in the D3 survey field.

We would like to use photometric redshifts for objects fainter than
the $I_{AB}<24.0$ VVDS spectroscopic limit. We can define another
figure of merit, the percentage of objects with
$\sigma_{sp}(68\%)<0.15(1+zp)$, where $\sigma_{sp}(68\%)$ is the
$68\%$ photometric redshift error bar.  This is plotted in Figure 15
in \citeauthor{2006A&A...457..841I} and gives an indication of how
good the photometric redshifts are beyond the spectroscopic limit. In
the interval $0.2<z_p<1.5$, this is always better than $80\%$ for all
four CFHTLS deep fields even as faint as $i'<24.5$.

\subsection{Computing absolute magnitudes and types}
\label{sec:comp-absol-magn}

We measure the absolute magnitude of each galaxy in $UBVRI$ standard
bands ($U$ Bessel, $B$ and $V$ Johnson, $R$ and $I$ Cousins). Using the
photometric redshift, the associated best-fit template and the observed
apparent magnitude in one given band, we can directly measure the
$k$-correction and the absolute magnitude in any rest-frame band. Since
at high redshifts the $k$-correction depends strongly on the galaxy
spectral energy distribution it is the main source of systematic error
in determining absolute magnitudes. To minimise $k$-correction
uncertainties, we derive the rest-frame luminosity at $\lambda$ using
the object's apparent magnitude closer to $\lambda \times (1+z)$. We
use either the $r'$, $i'$ or $z'$ observed apparent magnitudes
according to the redshift of the galaxy.  The procedure is described in
the Appendix A of \cite{2005A&A...439..863I} where it shown that this
method greatly reduces the dependency of the $k$-corrections on galaxy
templates.

\begin{figure}
\resizebox{\hsize}{!}{\includegraphics{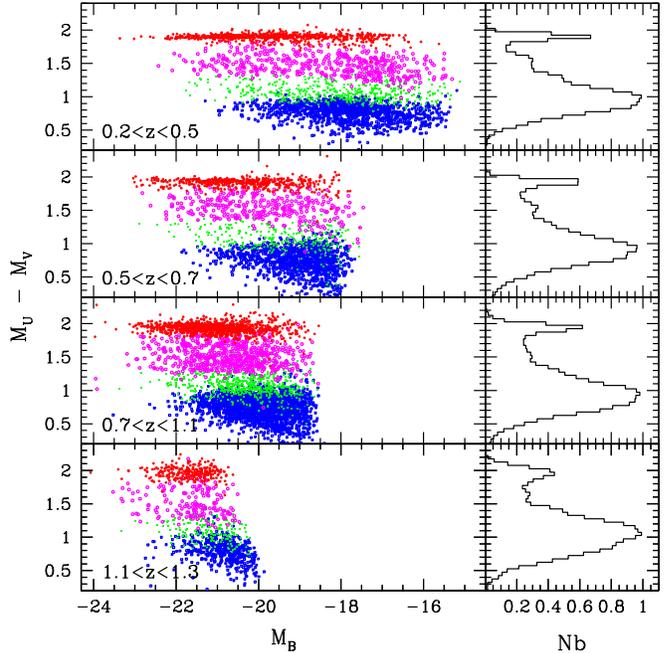}}
\caption{Rest frame $M_U-M_V$ colour as a function of B-band absolute
  magnitude for the D1 field. Each panel from top to bottom
  corresponds to the redshift bins used in this paper. The points show
  the four different best-fitting spectral types. In the colour
    version of this figure (available electronically) red, magnenta,
    green and blue points correspond to Coleman et al. Ell, Sbc, Scd,
    and Irr templates. The right-hand panels show the colour
  distribution for each redshift slice.}
\label{fig:absmag_bimod}
\end{figure}

Galaxies have been classified using multi-colour information in a
similar fashion to other works in the literature \citep[for
example][]{1999ApJ...518..533L,2003A&A...401...73W,2006A&A...455..879Z}.
For each galaxy the rest-frame colours were matched with four
templates from \cite{1980ApJS...43..393C} (hereafter referred to
``Coleman, Wu and Weedman'' or ``CWW'' templates).  These four
templates have been optimised using the VVDS spectroscopic redshifts,
as described in \citeauthor{2006A&A...457..841I}, and are presented in
Fig.2 of this work. Galaxies have been divided in four types,
corresponding to the optimised E/S0 template (type one), early spiral
template (type two), late spiral template (type three) and irregular
template (type four).  Type four includes also starburst galaxies.
Note that, in order to avoid introducing dependencies on any
particular model of galaxy evolution, we did not apply templates
corrections aimed at accounting for colour evolution as a function of
redshift.

We show in Figure~\ref{fig:absmag_bimod} the rest-frame colour
distribution of the galaxies for each type. Type one galaxies comprise
most of the galaxies of the red peak of the bimodal colour
distribution. The other types are distributed in the blue peak.
Galaxies become smoothly bluer from type one to type four respectively.

\section{Measuring galaxy clustering}
\label{sec:meas-galaxy-clust}

\subsection{Introduction}
\label{sec:introduction-m}
There are two principal approaches which may be used to measure the
clustering of objects with photometric redshifts. One is simply to
isolate galaxies in a certain redshift range using photometric
redshifts, and then to compute the projected correlation function
$w(\theta)$ for galaxies in this slice, as has long been done for
magnitude-limited samples. However, the additional information
provided by photometric redshifts on the \textit{bulk} properties of
our slice (its redshift distribution) allows us to use the Limber's
equation \citep{1953Apj...117..134} to invert the projected
correlation function and recover spatial correlations at the effective
redshift of the slice.  These computations are easy to perform and are
relatively insensitive to systematic errors in the photometric
redshifts as one just integrates over all galaxies in a given redshift
slice; it has already been used extensively in smaller surveys and is
usually the method of choice when only small numbers of galaxies or
poorer-quality photometric redshifts are available, and has been used
extensively over the past few years (see, for example
\cite{2001A&A...376..825D} or \cite{1999MNRAS.310..540A}).  It has the
disadvantage that it provides only limited information on the
\textit{shape} of the angular correlation function as one measures a
correlation function integrated over a given redshift slice.

A second approach is to decompose the redshift of each galaxy into
it's components perpendicular($r_p$) and parallel ($\pi$) to the
observer's line of sight, and then to compute a full two-dimensional
correlation function $\xi(r_p,\pi)$ based on pair counts of galaxies
in both directions.  Finally, one computes the sum of this clustering
amplitude in the direction parallel to the line of sight, $w_p$. In
spectroscopic surveys, this has the advantage of removing the effect
of redshift-space distortions caused by infall onto bound
structures. This technique has been used successfully for many
spectroscopic redshift surveys
\citep{1983ApJ...267..465D,1996ApJ...461..534L,2005ApJ...630....1Z}
and some attempts have been made to apply it to samples with
lower-accuracy photometric redshifts \citep{2006A&A...457..145P}. It
has the advantage that it can provide direct information on the shape
of the correlation function but this comes at the price of much
greater sensitivity to systematic errors in the photometric redshifts
(for example, integration over a much larger range in redshift space
is necessary).  We plan an analysis using this technique in a
forthcoming article, but in this paper we adopt a conservative
approach, as we are primarily interested in the overall clustering
properties of our galaxy samples.

\subsection{Projected angular clustering}
\label{sec:proj-angul-clust}

We first use our photometric redshift catalogue to produce a galaxy
sample generated using a given selection criterion, for example either
by absolute or apparent magnitude, redshift or type. This same
selection criterion is applied to catalogues for all four fields. 

From these masked catalogues of object positions, we measure
$w(\theta)$, the projected angular correlation function, using the
standard \citet{1993ApJ...412...64L} estimator:

\begin{equation}
w ( \theta) ={\mbox{DD} - 2\mbox{DR} + \mbox{RR}\over \mbox{RR}}
\label{eq:1.ls}
\end{equation}

where $DD$, $DR$ and $RR$ are the number of data--data, data--random
and random--random pairs with separations between $\theta$ and
$\theta+\delta\theta$. These pair counts are appropriately normalised;
we typically generate random catalogues with ten times higher numbers
of random points than input galaxies.

An important point to consider is that, of course, the precision of
our photometric redshifts are limited. In a given redshift interval,
$z_1<z<z_2$ it is certainly possible that a given galaxy may be in
fact outside this range. To account for this, we employ a weighted
estimator of $w(\theta)$, as suggested by
\cite{2002MNRAS.329..355A}. In this scheme we weight each galaxy by
the fraction of the galaxy's probability distribution function
enclosed by the interval $z_1<z<z_2$.

In this case for each pair we must now compute

\begin{equation}
DD=\sum_{i,j}^N{P^i}
{P^j}; DR=\sum_{i,j}^{N_d,N_r}{P^i}
\end{equation}

where $P^{i,j}$ is the integral of object's probability distribution
function in the redshift interval $z_1<z<z_2$. The total 'effective'
number of galaxies becomes 

\begin{equation}
N_{\rm eff}=\sum_{i=1}^{i=N_{gal}}P^i
\end{equation}

We may then fit $w(\theta)=A_w\theta^{-\delta}$ to find the
best-fitting amplitudes and slopes $A_w$ and $\delta$, after
first correcting for the ``integral constraint'' which arises because
the mean density is estimated by the survey itself,

\begin{equation}
C = {1 \over {\Omega^2}} \int \int \omega(\theta) d\Omega_1 d\Omega_2
\label{eq:2.iq}
\end{equation}

where $\Omega$ is the total area subtended by our survey. We compute C
by numerical integration of Equation~(\ref{eq:2.iq}), discounting
pairs closer than 1'', corresponding to the resolution limit of our
data. The integral constraint correction is subtracted from the power
law. We always fit our data on scales where theintegral constraint
correction is negligible.

We compute $w$ in a series of logarithmically spaced bins from
$\log(\theta)=-3$ to $\log(\theta)=-0.2$ with
$\delta\log(\theta)=0.2$, where $\theta$ is in degrees.  In the
following sections we will consider both measurements derived from
each field separately (using that field's redshift distribution) and
also measurements constructed from the sum of pairs over all fields
(in which case we use a combined, weighted redshift distribution
derived from all fields).

\subsection{Derivation of spatial quantities}
\label{sec:deriv-spat-quant}

We can associate each value of $A_w$ and $\delta$ with a
corresponding comoving correlation length, $r_0$ by making use of the
relativistic Limber equation \citep{P80}. For further discussion of
this method see, for example \cite{2001A&A...376..825D} or
\cite{1999MNRAS.310..540A}. If we assume that the spatial correlation
function can be expressed as $\xi(r)=(r/r_0)^{-\gamma}$ (where
$\gamma=1-\delta$ and $\Gamma$ is the incomplete Gamma function) 

\begin{equation}
w(\theta)={\sqrt\pi\Gamma((\gamma-1)/2)\over{\Gamma(\gamma/2)}}
{\int g(z)(dN/dz)^2r_0(z)^\gamma dz\over[\int(dN/dz)dz]^2}\theta^{1-\gamma}
\end{equation}

where $dN/dz$ represents the redshift distribution and $\theta$ is the
angular separation on the sky. Here g(z) depends only on cosmology and
is given by:

\begin{equation}
g(z)=(dx/dz)^{-1}x^{1-\gamma} F(x)
\end{equation}

with the metric defining $x$ and $F(x)$:
\begin{equation}
ds^2 = c^2dt^2 - a^2[dx/F(x)^2 + x^2 (\sin^2\theta d\phi^2)]
\end{equation}

Defining 

\begin{equation}
r_0^\gamma(z_{\rm eff})={\int^{z_2}_{z_1} g(z)(dN/dz)^2r_0(z)^\gamma
  dz \over\int_{z_1}^{z_2}g(z)(dN/dz)^2dz}
\label{eq:r0eff}
\end{equation}

and assuming that the correlation amplitude is constant over our
redshift interval $z_1<z<z_2$ and that $w(\theta) =
A_w\theta^{1-\gamma}$ we can write:

\begin{equation}
A_w=r_0^\gamma(z_{\rm eff})\sqrt\pi{\Gamma((\gamma-1)/2)\over{\Gamma(\gamma/2)}}{\int^{z_2}_{z_1} g(z)(dN/dz)^2
  dz \over[\int_{z_1}^{z_2}(dN/dz)dz]^2}
\end{equation}

Thus, given a measurement $A_w$ of the correlation function amplitude
for the redshift slice under consideration and a knowledge of that
slice's redshift distribution we can derive $r_0(z_{\rm eff})$. 

Given that we have four separate fields, we may derive $r_0$ from
a ``global'' $w(\theta)$ derived from the sum of pairs over all fields
and computed using the average, weighted $dn/dz$. Alternatively, we
can compute $r_0$ for each field using that field's redshift
distribution and individual correlation function amplitudes; the final
value of $r_0$ is then calculated simply as the average over all four
fields. We find that, in general, these methods agree for most
samples. However, in some cases the error bars are larger with the
field-to-field measurements. This is discussed in more detail in the
following section.

\subsection{Error estimates on $w(\theta)$}
\label{sec:error-estim-wtheta}
We have investigated two different approaches to estimate the errors
on our measurements of $r_0$. As described above, for each of our four
fields, we compute $r_0$ from a fit to a measurement of $w(\theta)$
and the redshift distribution \textit{for that field}. We then
compute the variance in $r_0$ and $\gamma$ over all fields. 

In the second method, for each sample, we compute the sum of the pairs
over all fields at each angular bin. The error bar at each angular bin
is then calculated from the variance in $w(\theta)$ over all
fields. If $w_{av}$ is the mean correlation function then $w_i$ is
the correlation function for each field, then the error over the $n$
fields of the CFHTLS is given as

\begin{equation}
\sigma^2={1\over(n-1)} {\sum_{i=1}^n(w_{av}-w_i)^2}
\label{eq:variance}
\end{equation}

where $n=4$ for the CFHTLS (note that $w_{av}$ is only used in this
computation and not in any other part of the analysis). 

A ``global'' correlation length is determined using the average redshift
distribution over four fields. In this case, the error in $r_0$ is
computed from the error in the best fitting $A_w$. This error is, in
turn, computed from the covariance matrix derived using the
Levenberg-Marquardt non-linear fitting routine, as presented in
Numerical Recipes \cite{NR}.

We find that in most cases the error bars in $r_0$ estimated by these
two different methods are consistent. However at lower redshifts
ranges ($0.2<z<0.6$), where the numbers of galaxies is smaller,
field-to-field dispersion is higher than the global errors.  Further
investigations reveal this is due to the presence of a single field
(d2) which has anomalously higher correlation lengths. Properties of
this field are discussed in detail in
\cite{2007ApJS..172..314M}. This is undoubtedly due to the presence
of very large structures at $z\sim0.3$ and $z\sim0.7$ in this field We
believe that our 'global' correlation function provides a more robust
estimate of the total error and we adopt this measurement for the
remainder of the paper.

\section{Results}
\label{sec:results}

\subsection{Magnitude limited samples}
\label{sec:magn-limit-sampl}

\begin{figure}
\resizebox{\hsize}{!}{\includegraphics{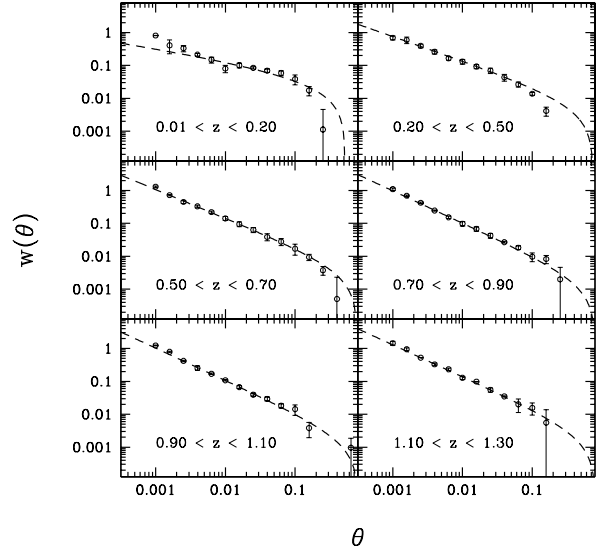}}
\caption{The amplitude of the angular correlation $w$ as a function of
  angular separation $\theta$ (in degrees) for $17.5<i'<24$
  galaxies selected in the four deep fields of the CFHTLS in a range
  of redshift slices.  The error bars correspond to the amplitude of
  the field-to-field variance over all fields. The dashed line shows
  the best-fitting power law correlation function after the
    subtraction of the appropriate integral constraint.}
\label{fig:wtheta}
\end{figure}

\begin{figure}
\resizebox{\hsize}{!}{\includegraphics{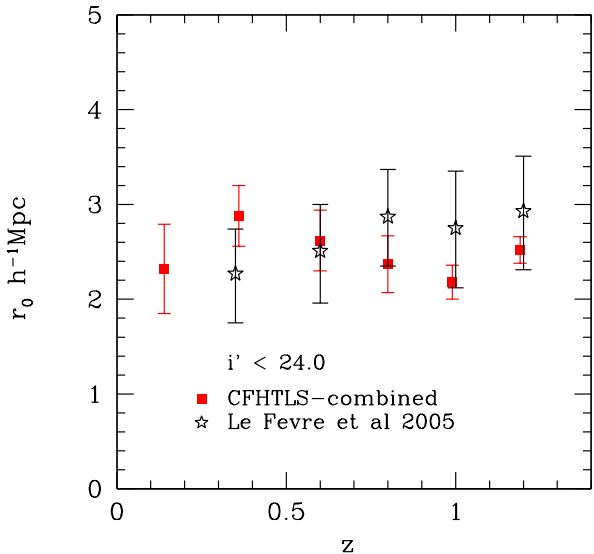}}

\caption{The comoving correlation length, $r_0$ as a function of
  redshift for the four combined CFHTLS fields (filled squares)
  compared to literature values (open symbols) computed for a galaxy
  sample limited at $i'<24.0$. For these fits, both $\gamma$ and $r_0$
  are free parameters. The error bars on these fits are computed from
  the field-to-field variance in each bin in $w$, as described in
  Section~\ref{sec:meas-galaxy-clust}. }

\label{fig:mz_vvds}
\end{figure}

\begin{figure}
\resizebox{\hsize}{!}{\includegraphics{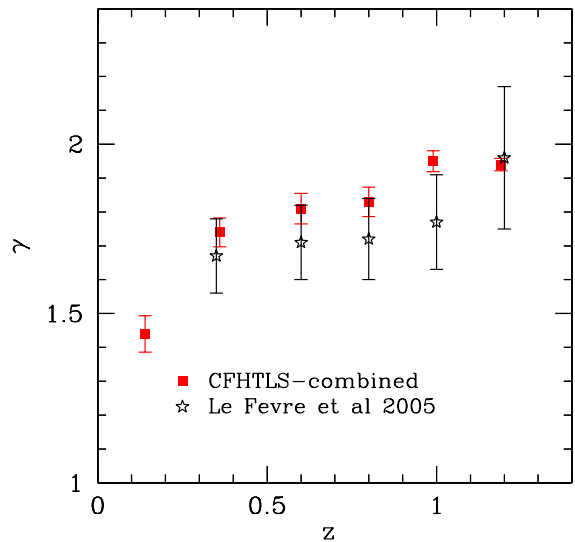}}
\caption{The best-fitting slope $\gamma$, as a function of sample
  median redshift for the four CFHTLS fields combined (filled squares)
  compared to literature measurements (open symbols) for a sample
  selected with $17.5<i'<24.0$ at each redshift slice.}
\label{fig:mz_slope}
\end{figure}

Can photometric redshifts be used to make reliable measurements of
galaxy clustering at $z\sim1$? In this Section we will construct a
sample similar to those already in the literature in order to address
this question. We consider galaxies selected by redshift and apparent
magnitude.  In each field we divide our catalogues into a series of
bins of width $\delta z=0.2$ over the range $0.2<z<1.2$. In each bin,
galaxies with $17.5<i_{AB}<24.0$ are selected to match the criterion
used by the VIMOS-VLT deep survey (VVDS) as presented in
\cite{2005A&A...439..877L} (we note that the Megacam instrumental $i'$
magnitudes are very close to the CFH12K instrumental $I$ magnitudes
used in \citeauthor{2005A&A...439..877L}).  Following the procedures
outlined in Section~\ref{sec:meas-galaxy-clust}, we measure the
weighted pair counts at each angular separation for each field and sum
them together.  Equation~(\ref{eq:1.ls}) is used to derive a
``global'' $w(\theta)$.  The error bar at each bin of angular
separation is computed from the variance of the \textit{individual}
measurements of $w(\theta)$ over all four fields, as described in
Equation~(\ref{eq:variance}).  These results are illustrated in
Figure~\ref{fig:wtheta}, which shows the amplitude of $w$ for all four
fields for a range of redshift slices. We note that in all redshift
slices except the lowest-redshift one, at intermediate scales,
$w(\theta)$ is well represented by a power law. In addition our error
bars are reassuringly small.  This global $w(\theta)$ is then fitted
with the usual power law, correcting for an integral constraint
corresponding to a total area of $3.2\deg^2$.

In calculating the correlation amplitude $r_0$ at the effective
redshift of each slice we use a redshift distribution derived from the
weighted, summed $dn/dz$ from each field.  Our results are displayed
in Figures~\ref{fig:mz_vvds} and \ref{fig:mz_slope}. They are
consistent with the measurements from the VVDS deep survey \citep[open
symbols;][]{2005A&A...439..877L}, which was based on a much smaller
sample of $\sim 7000$ spectroscopic redshifts. Our data does show some
evidence for a decline in the correlation amplitude strength in the
interval $0.5<z<1.1$, as well as a slightly higher slope, in contrast
with this earlier work. This effect will be discussed in more detail
in Section~\ref{sec:disc-results-comp}.

\subsection{Luminosity dependent clustering at $z\sim0.5$  and $z\sim1.0$}
\label{sec:lumin-deped-clust}

\begin{figure}
\resizebox{\hsize}{!}{\includegraphics{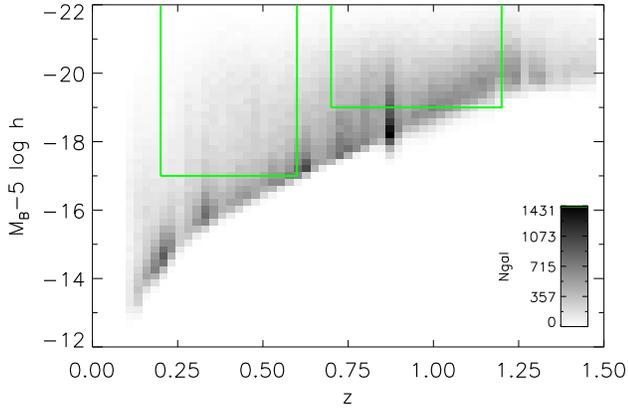}}
\caption{Gray-scaled histogram showing the distribution of galaxies as
  a function of absolute magnitude and redshift for four CFHTLS fields
  for all galaxy types and for an apparent magnitude limit of
  $i'<24.5$.}
\label{fig:mz_absmag_allgal}
\end{figure}

The main difficulty in interpreting Figures~\ref{fig:mz_vvds} and
\ref{fig:mz_slope} is that at each redshift slice the sample's median
absolute luminosity changes significantly as a consequence of the
selection in apparent magnitude. This is evident if one considers
Figure~\ref{fig:mz_absmag_allgal} which shows a two-dimensional image
of the objects distribution in the absolute magnitude-redshift plane.
Slices extracted at lower redshift are dominated by galaxies of
intrinsically low absolute luminosity.

\begin{figure}[!h]
\resizebox{\hsize}{!}{\includegraphics{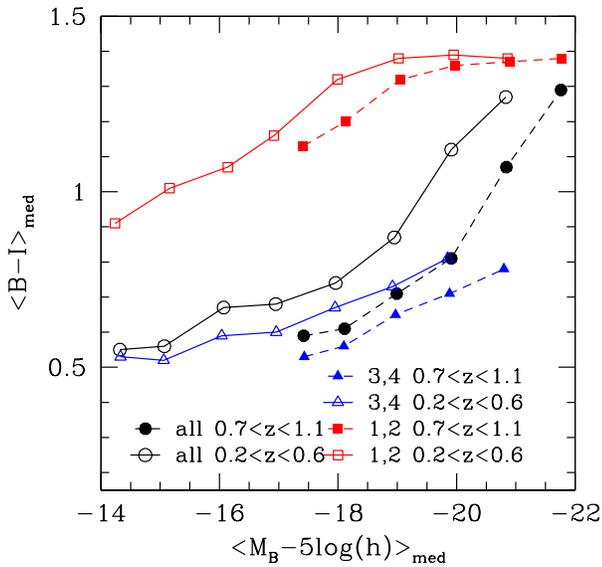}}
   \caption{Median rest-frame (B-I) colour versus median rest frame
     $B-$band absolute luminosity for samples at $0.2<z<0.6$ (open
     symbols, solid lines) and $0.7<z<1.1$ (filled symbols, dotted
     lines). For each redshift range we show all galaxy types
     (circles), types one and two (squares) and types three and four
     (triangles).}
\label{fig:absmag_colour}
\end{figure}

To investigate the dependence of galaxy clustering amplitude on
absolute luminosity, in this Section we extract samples \emph{in two
  fixed redshift intervals} selected from the
absolute-magnitude/luminosity plane. We consider galaxies between
$0.2<z<0.6$ and $0.7<z<1.1$. In each redshift range we select a
minimum absolute magnitude (shown by the solid lines in
Figure~\ref{fig:mz_absmag_allgal}) so that the median redshift of
galaxies selected in each slice of absolute luminosity is
approximately constant.  At each redshift interval we separate the
galaxy population into ``early'' (types one and two) and ``late''
(types three and four). We also consider samples comprising all galaxy
types. These samples are illustrated in
Figure~\ref{fig:absmag_colour}, where the median rest-frame Johnson
$(B-I)$ colour is plotted as a function of median rest-frame $M_B$
magnitude. In both high- and low- redshift slices, changes in the
absolute magnitude bin produces the largest changes in rest-frame
colours. It is also clear that galaxy populations become progressively
bluer at higher redshifts (dotted lines and filled symbols), at the
same bin in absolute magnitude.

\begin{figure*}
\centering
\includegraphics[width=17cm]{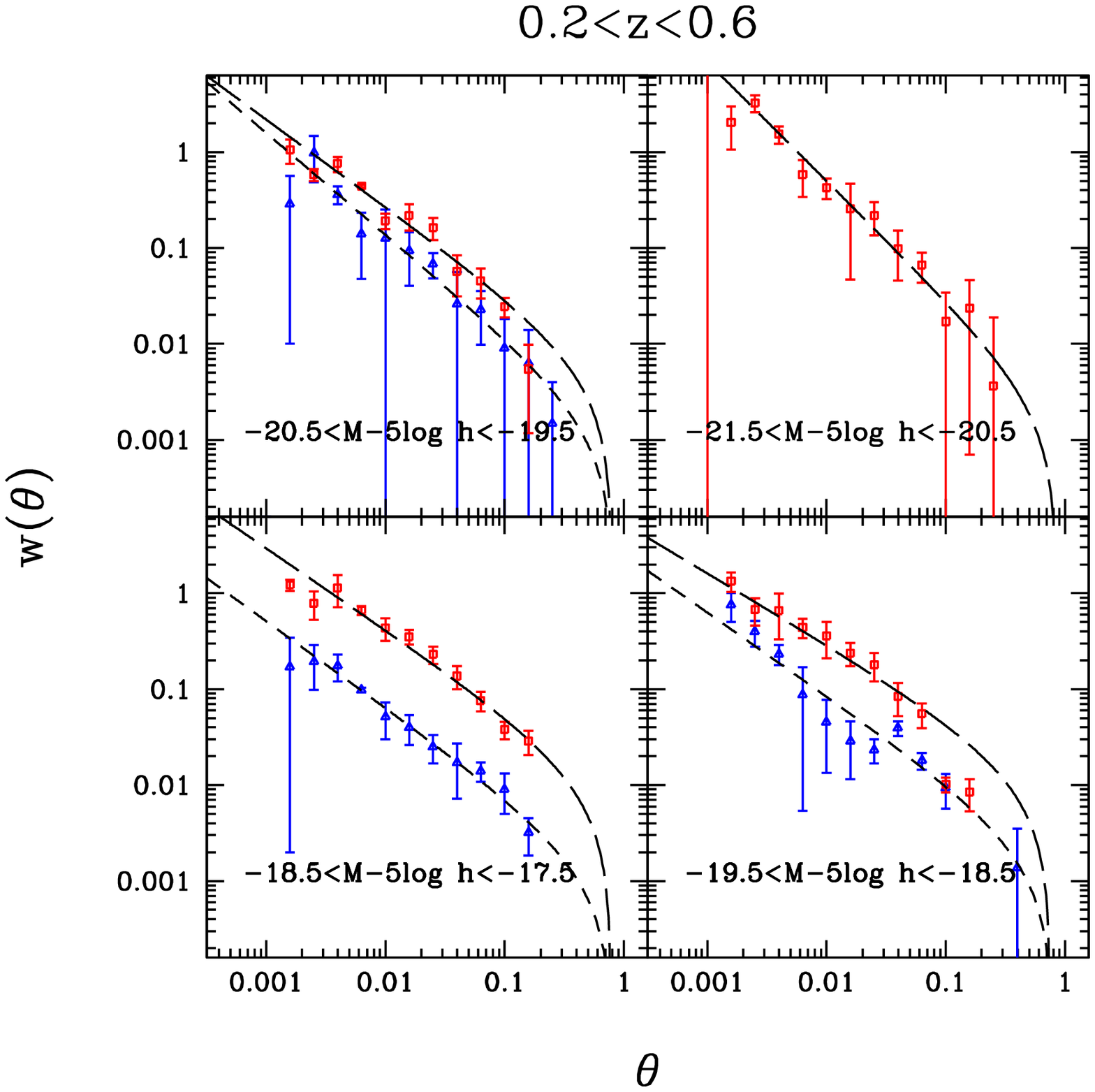}
\caption{The amplitude of the angular correlation $w$ as a function of
  angular separation $\theta$ (in degrees) for $i'<24.5$ galaxies in
  the redshift range $0.2<z<0.6$. Each panel shows galaxies selected
  in a different, independent, slice in absolute luminosity. Squares
  represent the early-type population (types 1 and 2) whereas
  triangles show the late-type population. The dashed and long-dashed
  lines shows the adopted best-fit. The error bars at each angular
  separation correspond to field-to-field variance measured over the
  four survey fields.}

\label{fig:wtheta_bytype}
\end{figure*}

\begin{figure*}
\centering
\includegraphics[width=17cm]{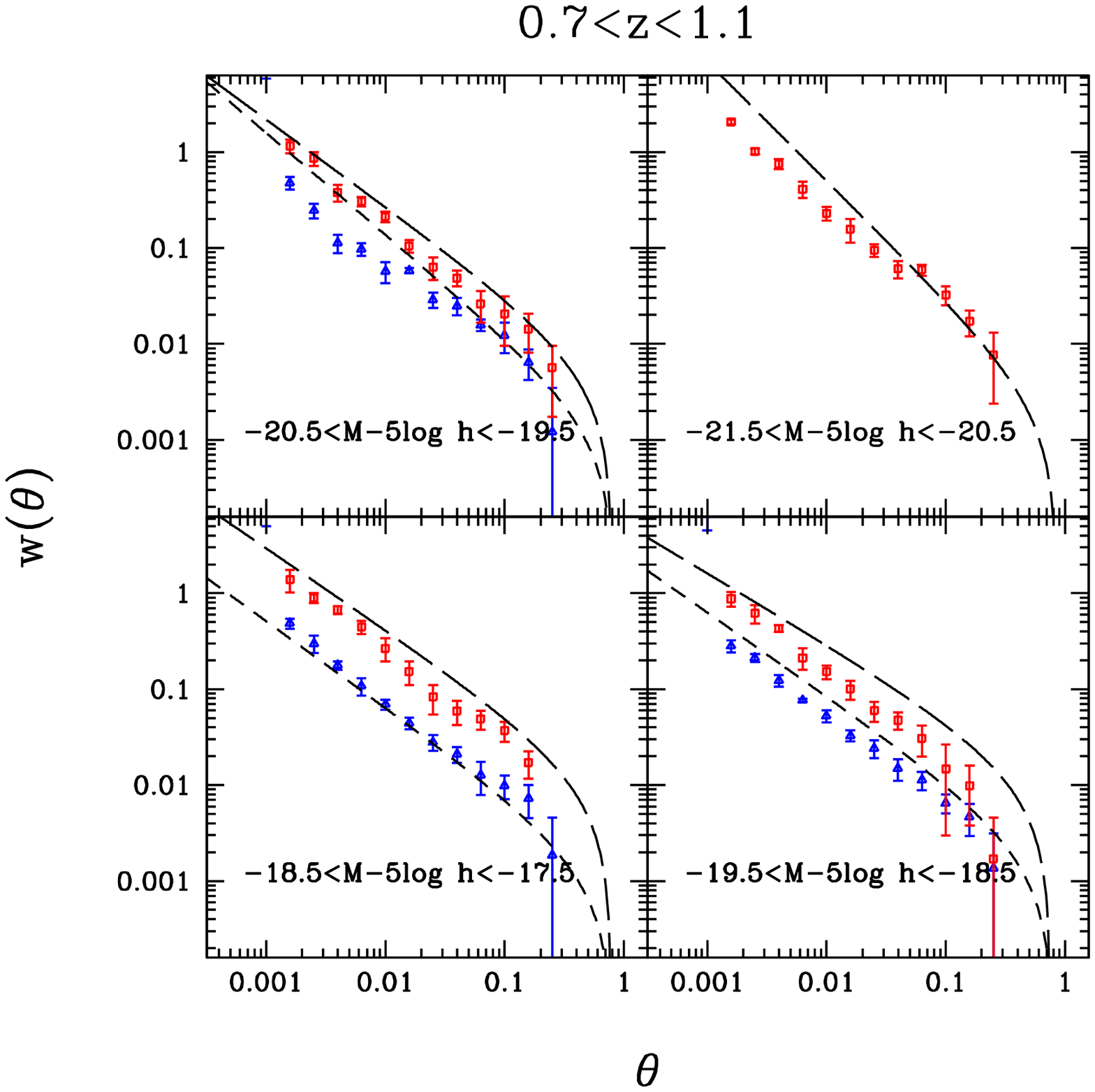}
\caption{Similar to Figure~\ref{fig:wtheta_bytype}: the amplitude of
  the angular correlation $w$ as a function of angular separation
  $\theta$ (in degrees) for red and blue $i'<24.5$ galaxies in the
  redshift range $0.7<z<1.1$. The dashed and long-dashed lines shows
  the adopted best-fit for the \emph{same bin} in absolute luminosity
  in the lower redshift range $0.2<z<0.6$. The error bars at each
  angular separation correspond to field-to-field variance measured
  over the four survey fields.}
\label{fig:wtheta_bytype_hz}
\end{figure*}

Our type-selected correlation functions for galaxies in the redshift
range $0.2<z<0.6$ are displayed in Figure~\ref{fig:wtheta_bytype}, and
for the higher redshift range in
Figure~\ref{fig:wtheta_bytype_hz}. These plots show the amplitude of
the angular correlation function $w$ as a function of angular
separation, $\theta$, for different slices of absolute magnitude. In
each panel we show correlation functions measured for the red and blue
(early and late) populations. The size of the error bars at each
angular separation corresponds to the amplitude of the field-to-field
cosmic variance computed over the four fields. The long-dashed and
dashed lines show the fitted amplitudes for the red and blue
populations respectively. For the higher redshift bin, we superimpose
the fitted amplitudes at the same bin in absolute luminosity at the
lower redshift interval. At all redshifts and absolute luminosity
ranges, galaxies with redder rest-frame colours are more clustered
than their bluer counterparts.

\begin{figure}
\resizebox{\hsize}{!}{\includegraphics{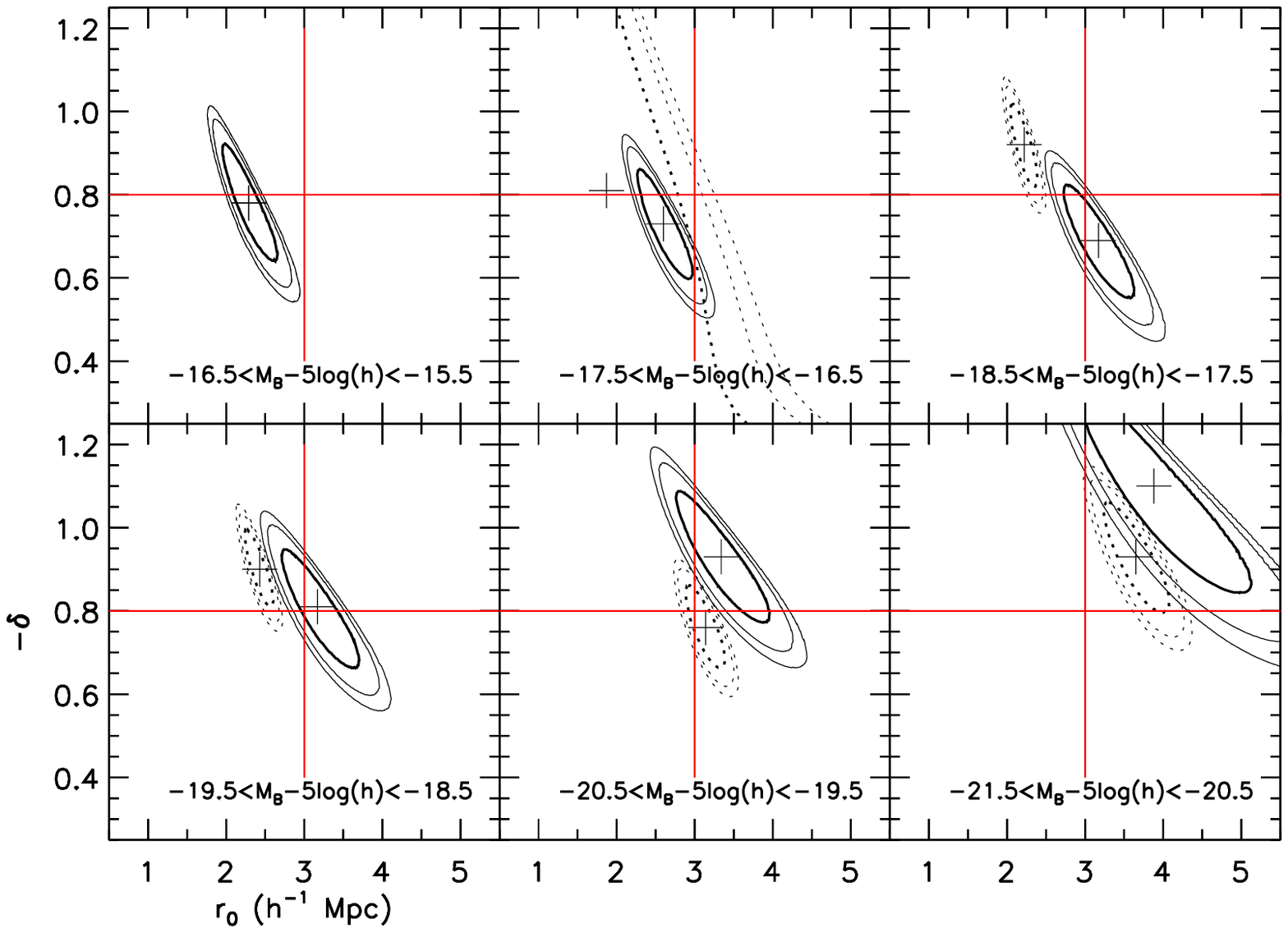}}
\caption{The comoving correlation length $r_0$ and slope
  $\delta=1-\gamma$ as a function of median absolute luminosity and
  redshift for low-redshift ($0.2<z<0.6$; solid lines) and high
  redshift ($0.7<z<1.1$; dotted lines) populations for four fields of
  the CFHTLS-deep. Each panel shows contours of constant chi-squared
  values for $1,2,$ and $3\sigma$ confidence levels with the plus
  symbol marking the minimum chi-squared value. For convenience, lines
  of constant slope and comoving correlation length are plotted.}
\label{fig:mz_absmag}
\end{figure}

Because of the strong covariance between $\gamma$ and $r_0$ it is
useful to consider contours of constant $\chi^2$ at each slice in
absolute magnitude. This is shown in Figure~\ref{fig:mz_absmag}. The
vertical and horizontal lines mark an arbitrary reference point. Solid
lines indicate galaxies with $0.2<z<0.6$ and dotted lines those with
$0.7<z<1.1$. From these plots we see a gradual increase in comoving
correlation length as a function of absolute rest-frame luminosity. We
also see some evidence for a slight decrease in the comoving
correlation length at given fixed absolute luminosity between higher
and lower redshift slices.

\begin{figure}
\resizebox{\hsize}{!}{\includegraphics{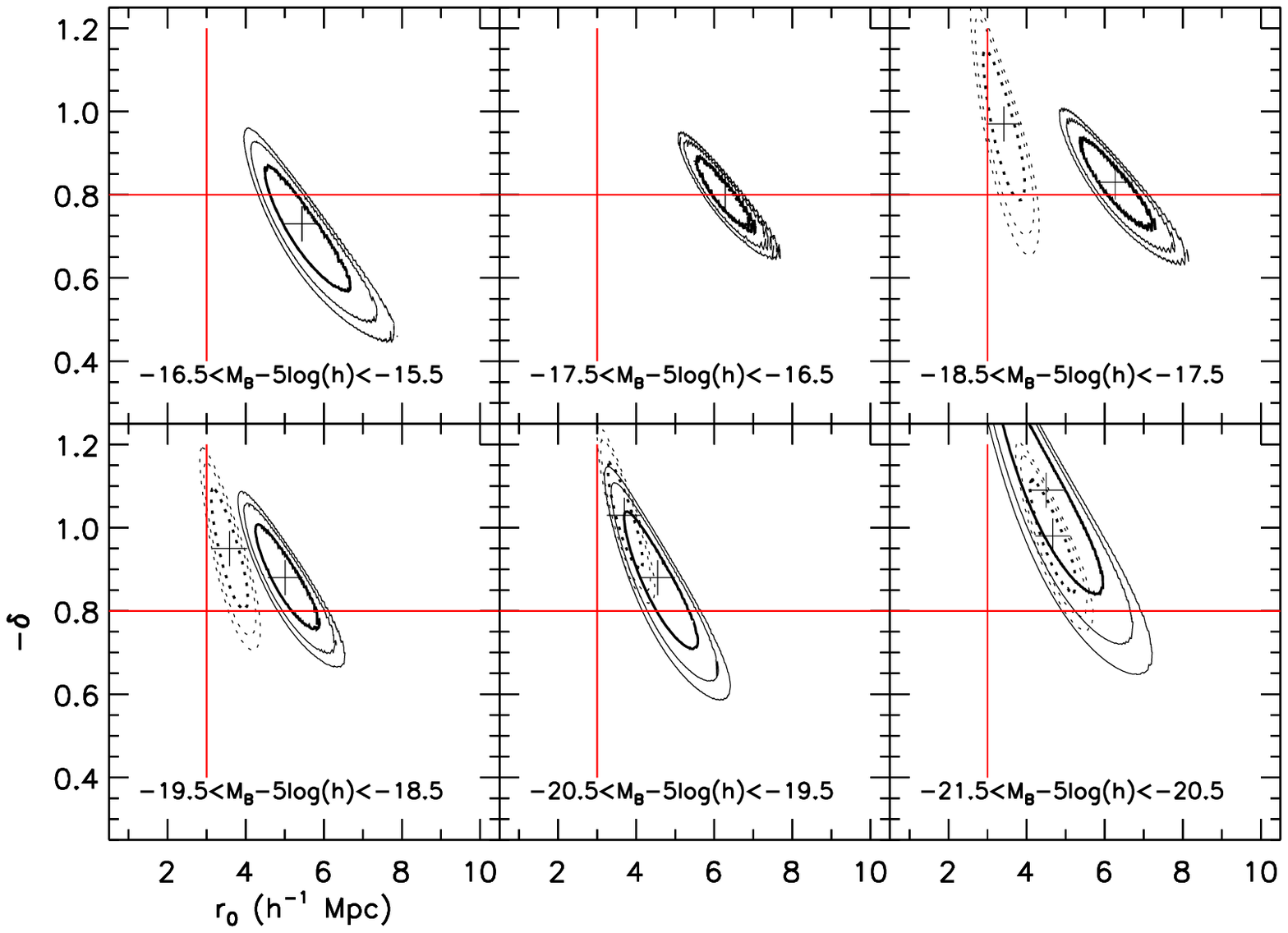}}
\caption{The comoving correlation length $r_0$ and slope $\delta=
  1-\gamma$ as a functions of absolute luminosity and redshift for
  early-type galaxies for the four fields of the CFHTLS.  Each panel
  shows contours of constant chi-squared values for $1$,$2$, and
  $3\sigma$ confidence levels with the plus symbol marking the
  minimum. The solid and dotted lines shows galaxies in the redshift
  bin $0.2<z<0.6$ and $0.7<z<1.1$ respectively for the same range of
  absolute luminosities.  We note that in the
  $-18.5<M_B-5\log(h)<-17.5$ and $-19.5<M_B-5\log(h)<-18.5$ luminosity
  ranges the clustering amplitude of early type galaxies at different
  redshifts are well separated.}
\label{fig:mz_absmag_bytype_early}
\end{figure}

\begin{figure}
\resizebox{\hsize}{!}{\includegraphics{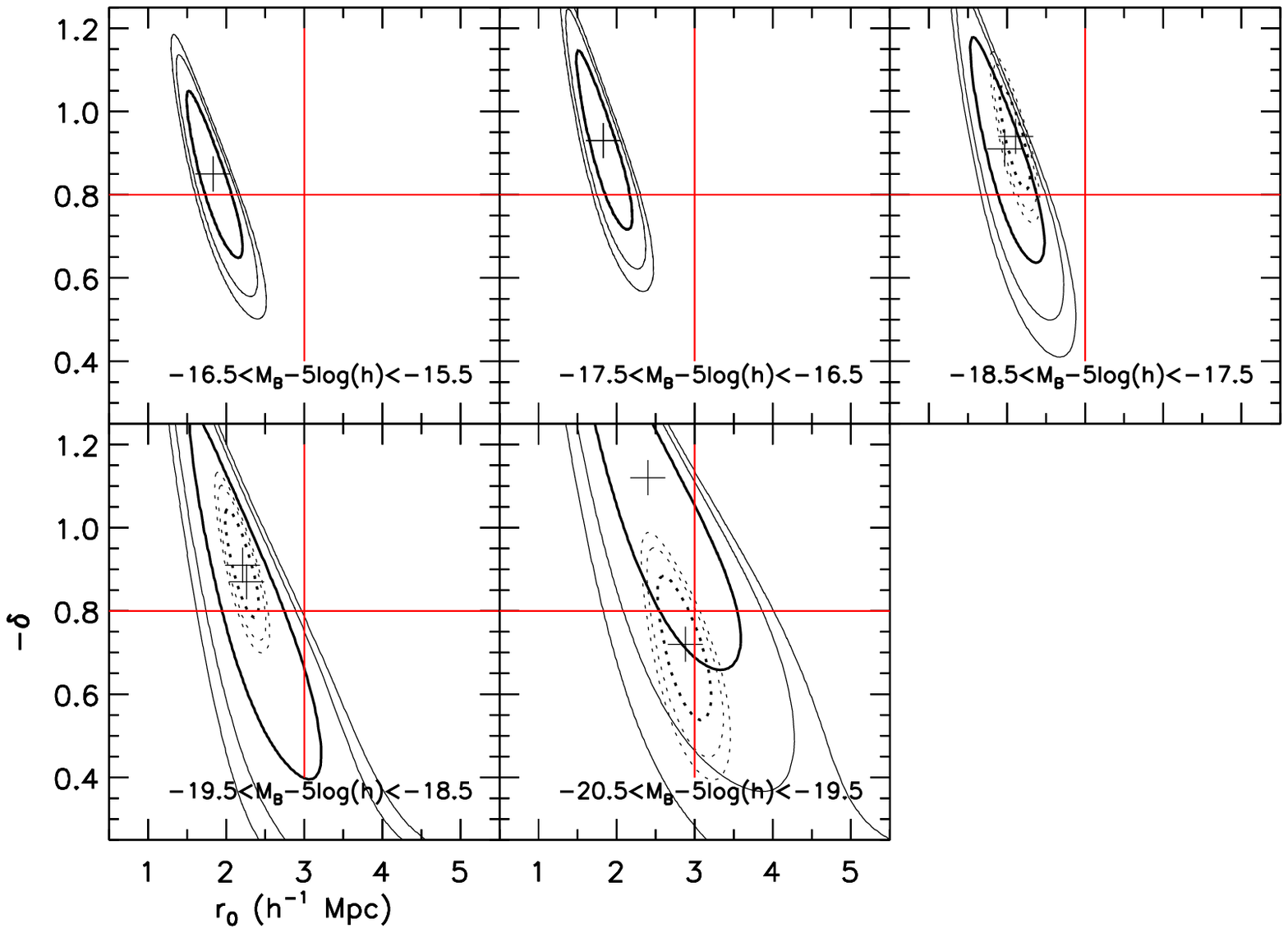}}
\caption{The comoving correlation length $r_0$ and slope
  $\delta=1-\gamma$ as a function of absolute luminosity and redshift
  for late-type populations for the four fields of the CFHTLS.  Each
  panel shows contours of constant chi-squared values for $1$,$2$, and
  $3\sigma$ confidence levels with the plus symbol marking the
  minimum.  The solid and dotted lines shows galaxies in the
  redshift bin $0.2<z<0.6$ and $0.7<z<1.1$ respectively for the same
  range of absolute luminosities.}
\label{fig:mz_absmag_bytype_late}
\end{figure}

Figures~\ref{fig:mz_absmag_bytype_early} and
\ref{fig:mz_absmag_bytype_late} show the comoving correlation function
$\chi^2$ contours for red and blue populations in both redshift
ranges; and correspond to the angular correlation functions presented
in Figures~\ref{fig:wtheta_bytype} and
Figures~\ref{fig:wtheta_bytype_hz}. Figure~\ref{fig:mz_absmag} shows
the comoving correlation length for the full galaxy population in both
redshift ranges.

\begin{figure}
\resizebox{\hsize}{!}{\includegraphics{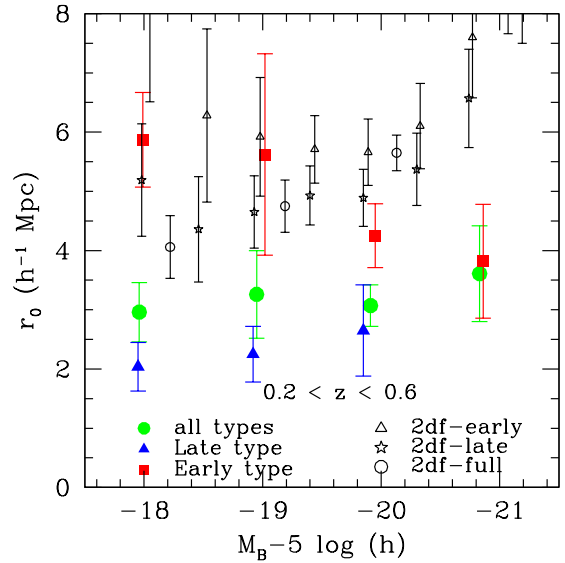}}
\caption{The comoving correlation length $r_0$ as a function of median
  absolute luminosity and type for objects in the redshift range
  $0.2<z<0.6$. Filled circles show the full galaxy population. In
  addition to type selection, the galaxy sample is selected in
  one-magnitude bins of absolute luminosity. Triangles and squares
  represent the late and early-type populations respectively.}
\label{fig:mz_absmag_lz}
\end{figure}

\begin{figure}
\resizebox{\hsize}{!}{\includegraphics{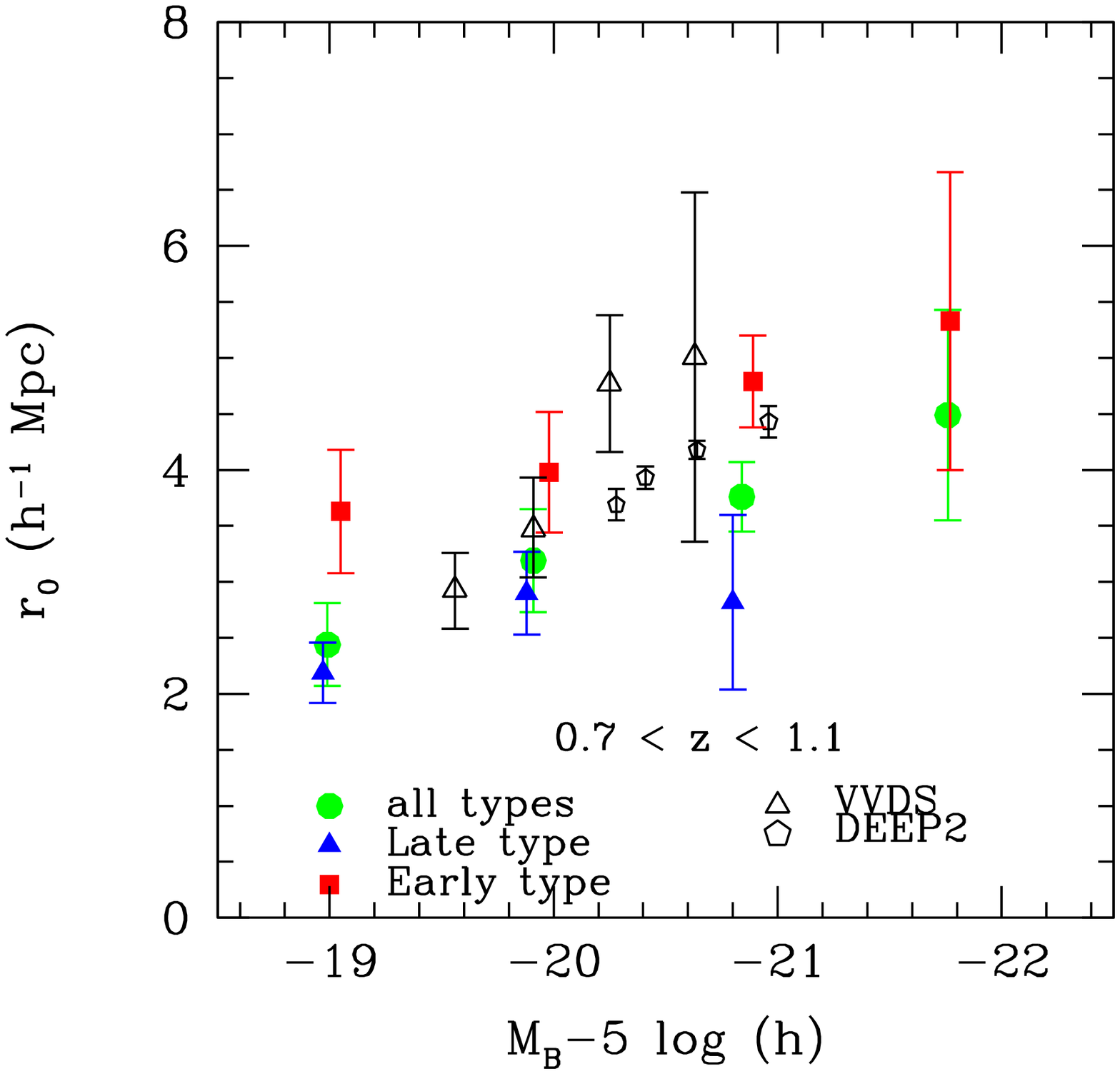}}
\caption{Similar to Figure~\ref{fig:mz_absmag_lz}; here we show
  measurements from our $0.6<z<1.1$ sample. Points from the literature
  from the measurements made by the DEEP2 and VVDS surveys.}
\label{fig:mz_absmag_hz}
\end{figure}

The results presented in this Section are summarised in
Figures~\ref{fig:mz_absmag_lz}, \ref{fig:mz_absmag_hz} and in
Tables~\ref{tab:r0_results_lz} and \ref{tab:r0_results_hz}. These
figures show the best-fitting correlation amplitude as a function of
absolute luminosity for the three different samples (early, late and
the full galaxy population) in the two redshift ranges ($0.2<z<0.6$
and $0.7<z<1.1$) considered in this Section.

Considering these plots, several features are apparent. Firstly, at
all absolute magnitude slices and in both redshift ranges, early-type
galaxies are \textit{always} more strongly clustered (higher values of
$r_0$) than late-type galaxies.  Secondly, we note that clustering
amplitude for the late-type population is remarkably constant,
remaining fixed at $\sim2h^{-1}$~Mpc over a large range of absolute
magnitudes and redshifts. The behaviour of the early-type population
is more complicated. For the $0.2<z<0.6$ bin, we some evidence that as
the median luminosity increases, the clustering amplitude of this
population \textit{decreases}, from around $\sim 6 h^{-1}$~Mpc for the
faintest bins, to $\sim 5h^{-1}$~Mpc. We note that the difference in
clustering amplitude between the early and late populations is smaller
for the higher-redshift bin. We also note that the clustering
amplitudes we derive for our blue and full-field galaxy populations
are considerably lower than those reported by
\cite{2002MNRAS.332..827N} at lower redshifts.

We carried out several test to verify the robustness of the higher
correlation amplitudes observed for the fainter red galaxy population
at $z\sim0.5$. Selecting galaxies with redder rest-frame colours
(classified as type one) in larger bins of absolute magnitude we also
measure higher clustering amplitudes for objects fainter than $M_B-5
\log h\sim-19$. We also note that the origin of the large error bar
for the bin at $M_B-5 \log h = -19$ is due to the presence of structures
in one of the four fields; interestingly, for the D1 field, for this
faint red population, the correlation function does not follow a
normal power-law shape. Our resulting error bars reflect this
behaviour, but it is clear that for certain galaxy populations, for
instance the bright elliptical population, simple power-law fits are
not appropriate.

In the redshift range $0.7<z<1.1$ we have compared our measurements of
pure luminosity dependent clustering (i.e., without type selection) to
works in the literature computed using smaller samples of
spectroscopic redshifts, namely \cite{2006A&A...451..409P} and
\cite{2006ApJ...644..671C}, shown in Figure~\ref{fig:mz_absmag_hz} as
the open triangles and open stars respectively. Their points should be
compared with the full circles derived from our measurements. In
general the agreement is acceptable, although for higher luminosity
bins, our amplitudes are below the measurements from the DEEP2 survey
(although it seems that the amplitude of their error bars is perhaps
underestimated).

\begin{figure}
\resizebox{\hsize}{!}{\includegraphics{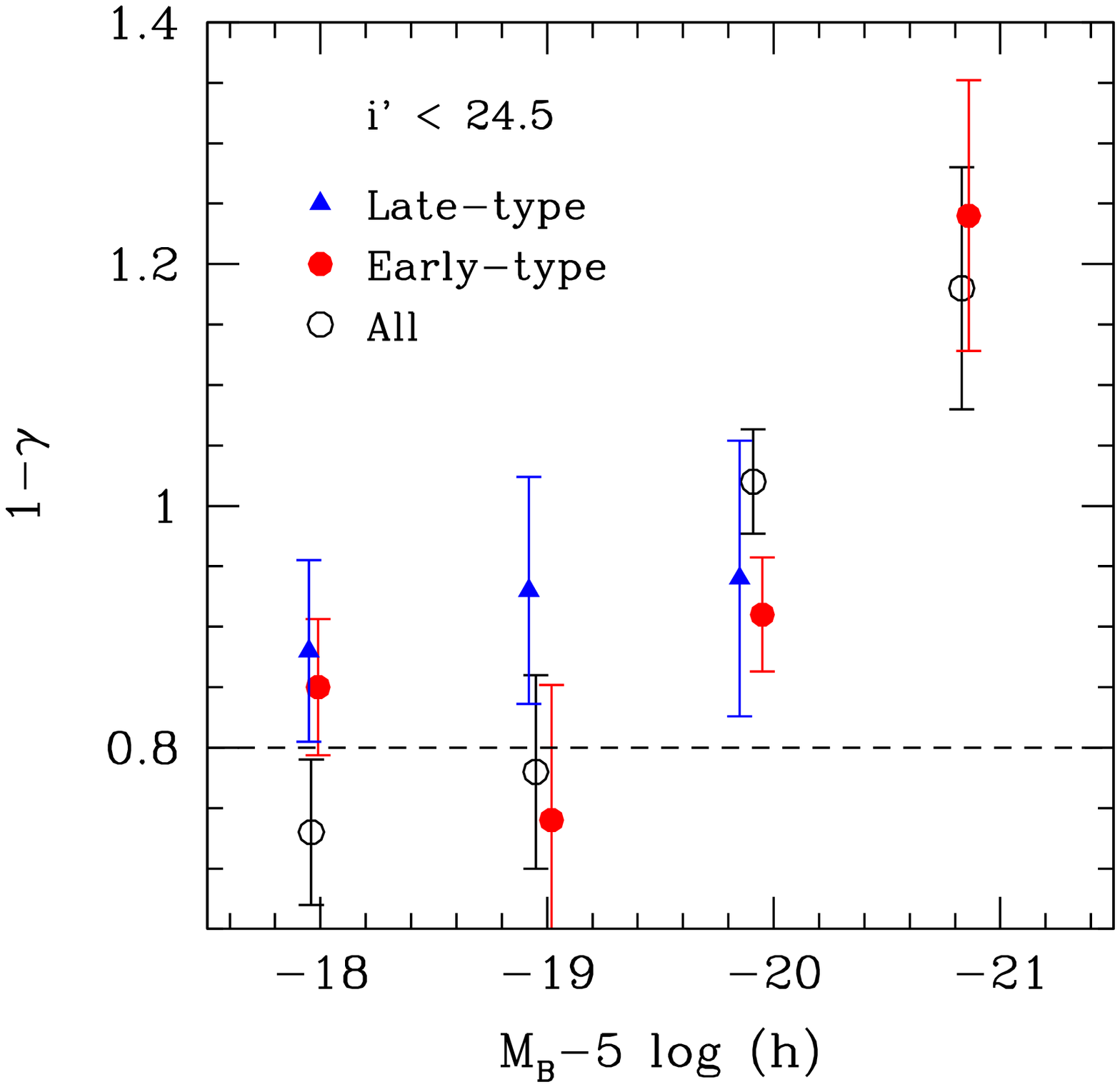}}
\caption{Best-fitting slope $1-\gamma$ of $w$ as a function of median absolute
  luminosity at $0.2<z<0.6$ for the early-type population (filled
  circles), late-type population (triangles) and full population (open
  circles). Slopes are plotted as a function of the median absolute
  magnitude in each slice of one magnitude in width.}
\label{fig:slope_lz}
\end{figure}

\begin{figure}
\resizebox{\hsize}{!}{\includegraphics{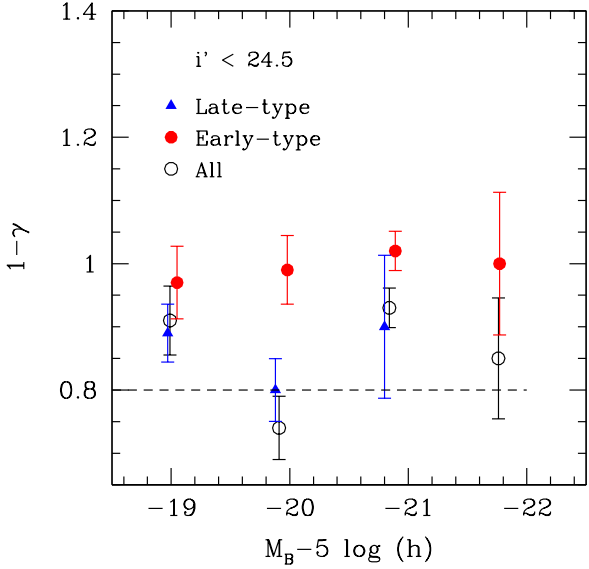}}
\caption{As in Figure~\ref{fig:mz_absmag_lz} but for the redshift
  range $0.7<z<1.1$. Slopes are plotted as a function of the median
  absolute magnitude in each bin of one magnitude in width.}
\label{fig:slope_hz}
\end{figure}

For all the plots previously shown in this Section we fitted
simultaneously for the slope and amplitude of the galaxy correlation
function. In Figure~\ref{fig:slope_lz} and Figure~\ref{fig:slope_hz}
we summarise our results from our low and high redshift
samples. Figure~\ref{fig:slope_hz} shows the slope $1-\gamma$ as a
function of absolute luminosity for early-type, late type, and
full-field populations at high redshifts. Figure~\ref{fig:slope_lz}
presents the results from the $0.2<z<0.6$ sample. Error bars are
computed from the field-to-field variance.

Interestingly, we find for the higher redshift bin ($0.7<z<1.1$) the
slope is relatively insensitive to absolute magnitude. However, at
lower redshifts, luminous red galaxies have a steeper correlation
function slope than fainter galaxies. A similar effect is observed in
the SDSS and two-degree field surveys at lower
redshifts \citep{2002MNRAS.332..827N}.

\begin{table}
\begin{tabular}{c c c c c c}
\hline\hline
$\langle M_B\rangle$ & $N_{\rm gals}$ & $\langle B-I\rangle$ &$z_{\rm eff}$ & $r_0$ & $\delta$\\ 
\hline
\\All& & & & \\
-18.0 & 15741 &  0.74 &  0.46 &$3.0\pm0.5$ & $0.7\pm0.06$\\
-18.9 & 11025 &  0.87 &  0.46 &$3.2\pm0.3$ & $0.9\pm0.06$\\
-19.9 & 6607 &  1.12 &  0.45 &$3.1\pm0.3$ & $1.0\pm0.04$\\
-20.8 & 2153 &  1.27 &  0.46 &$3.6\pm0.8$ & $1.2\pm0.10$\\
& & & & & \\
Red& & & & \\
-18.0 & 3704 &  1.32 &  0.42 &$5.9\pm0.8$ & $0.8\pm0.06$\\
-19.0 & 3957 &  1.38 &  0.43 &$5.6\pm1.7$ & $0.7\pm0.11$\\
-19.9 & 3656 &  1.39 &  0.44 &$4.2\pm0.5$ & $0.9\pm0.05$\\
-20.9 & 1567 &  1.38 &  0.45 &$3.8\pm1.0$ & $1.2\pm0.11$\\
& & & & & \\
Blue& & & & \\
-17.9 & 12037 &  0.67 &  0.47 &$2.0\pm0.4$ & $0.9\pm0.07$\\
-18.9 & 7068 &  0.73 &  0.47 &$2.2\pm0.5$ & $0.9\pm0.09$\\
-19.9 & 2951 &  0.81 &  0.47 &$2.6\pm0.8$ & $0.9\pm0.11$\\
\\
\end{tabular}
\caption{Low redshift sample ($0.2<z<0.6$). Columns show the median
  rest-frame $B-$band absolute luminosity, total number of galaxies over
the four fields, median absolute rest-frame B-I colour, effective
redshift, best fitting correlation length and slope.}
\label{tab:r0_results_lz}
\end{table}

\begin{table}
\centering
\begin{tabular}{c c c c c c}
\hline\hline
$\langle M_B \rangle$ & $N_{gals}$ & $\langle B-I\rangle$ &$z_{\rm eff}$ & $r_0$ & $\delta$\\ 
\hline
\\All& & & & \\
-19.0 & 40483 &  0.71 &  0.90 &$2.4\pm0.4$ & $0.9\pm0.05$\\
-19.9 & 25452 &  0.81 &  0.91 &$3.2\pm0.5$ & $0.7\pm0.05$\\
-20.8 & 9689 &  1.07 &  0.91 &$3.8\pm0.3$ & $0.9\pm0.03$\\
-21.8 & 1664 &  1.29 &  0.93 &$4.5\pm0.9$ & $0.8\pm0.10$\\
& & & & & \\
Red& & & & \\
-19.1 & 8639 &  1.32 &  0.88 &$3.6\pm0.6$ & $1.0\pm0.06$\\
-20.0 & 8859 &  1.36 &  0.90 &$4.0\pm0.5$ & $1.0\pm0.05$\\
-20.9 & 5511 &  1.37 &  0.91 &$4.8\pm0.4$ & $1.0\pm0.03$\\
-21.8 & 1280 &  1.38 &  0.93 &$5.3\pm1.3$ & $1.0\pm0.11$\\
& & & & & \\
Blue & & & & \\
-19.0 & 31844 &  0.65 &  0.91 &$2.2\pm0.3$ & $0.9\pm0.05$\\
-19.9 & 16593 &  0.71 &  0.91 &$2.9\pm0.4$ & $0.8\pm0.05$\\
-20.8 & 4178 &  0.78 &  0.91 &$2.8\pm0.8$ & $0.9\pm0.11$\\
\\
\end{tabular}
\caption{High redshift sample ($0.7<z<1.1$). Columns show the median
  rest-frame $B-$band absolute luminosity, total number of galaxies over
the four fields, median absolute rest-frame B-I colour, effective
redshift, best fitting correlation length and slope.}
\label{tab:r0_results_hz}
\end{table}

\subsection{Clustering of bright early type galaxies}
\label{sec:clust-bright-early}

\begin{figure}
\resizebox{\hsize}{!}{\includegraphics{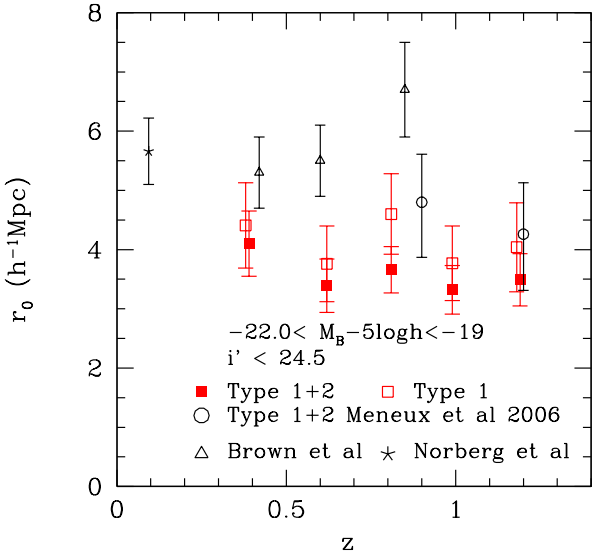}}
\caption{Clustering amplitude of luminous red galaxies. Open and
  filled squares show measurements for type one and type one and two
  combined galaxy samples. Other points show measurements from the
  literature for early type galaxies selected using a variety of
  methods.}
\label{fig:comparison}
\end{figure}

In Figure~\ref{fig:comparison} we plot the clustering amplitudes of
bright early-type galaxies in our survey; filled squares indicate
galaxies of types one and two, and open squares represent a pure type
one sample. As expected, the clustering amplitudes of the pure type
one population (with overall redder rest-frame colours) are higher
than the combined type one and two samples. (We have also measured the
clustering amplitude of the pure type four population and find that in
this case clustering amplitudes are lower than the combined sample of
types three and four.) Error bars are computed from the field-to-field
variance.

Several authors have presented clustering measurements as a function
of either absolute luminosity, type or redshift. For example,
\cite{2006A&A...452..387M} described measurements in the VVDS
spectroscopic redshift survey of the projected correlation function
for early- and late-type galaxies. Their galaxies are classified in
the same way as in this paper, using CWW templates.  However, in their
sample galaxies were selected by apparent magnitude; at $z\sim1$,
their rest frame luminosities are comparable to the brightest galaxies
in our sample. We compare these $z\sim1$ with our data; they are shown
as the open circles in Figure~\ref{fig:comparison}. Finally, the open
triangles represent measurements from \citet{2003ApJ...597..225B} who
measured clustering of red galaxies selected using three-band
photometric redshifts in the NOAO wide survey. Their results are above
ours by at least one or two standard deviations. In general we note
that our results are lower than literature measurements and speculate
that this could be the consequence of a slight loss of signal due our
use of photometric redshifts.  The broad trend seen in our
measurements is that the clustering amplitude of bright early-type
galaxies does not change with redshift.

\subsection{Redshift-dependent clustering}
\label{sec:volume-type-depend}

\begin{figure}
\resizebox{\hsize}{!}{\includegraphics{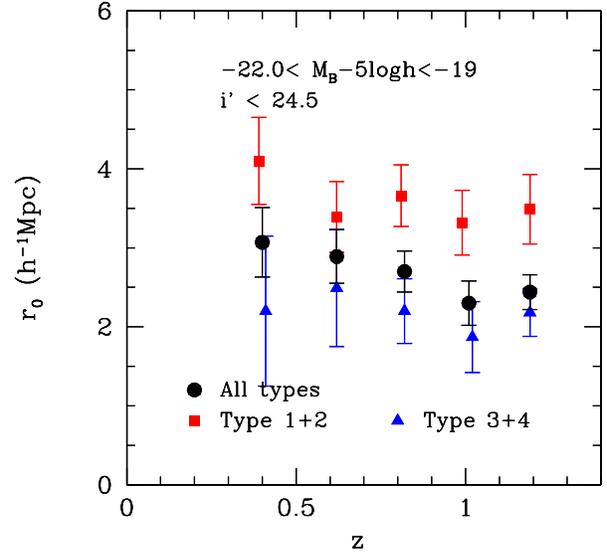}}
\caption{Redshift dependence of comoving galaxy correlation length
  $r_0$ for a series of volume limited samples for early types
  (squares), late types (triangles) and for the full sample (filled circles).}
\label{fig:mz_absmag_new}
\end{figure}

How does the comoving correlation length of the various galaxy
populations investigated here depend on redshift? Over the full range
redshift range ($0.2<z<1.1$), as is apparent from
Figure~\ref{fig:mz_absmag_allgal}, this measurement is only possible
for the brightest galaxies; at higher redshifts, intrinsically fainter
galaxies drop out of our survey. For each of the redshift bins used in
Figure~\ref{fig:wtheta} we selected galaxies with $-22.0 < M_B-5 \log
h<-19$ and measured $r_0$ and $\gamma$ as above.  The derived
amplitudes for this sample are shown as the stars in
Figure~\ref{fig:mz_absmag_new}. We also selected at each redshift bin
early type galaxies (types 1 and 2) and late type galaxies (types 3
and 4); these are represented by squares and triangles in
Figure~\ref{fig:mz_absmag_new}. We find that the median absolute
magnitude at each bin is $M_B-5\log h \sim -19.6$ for the early types
and $M_B-5\log h \sim19.4$ for the late types. The full galaxy
population has an absolute magnitude of $M_B-5\log h \sim-19.5$. From
$0.4<z<1.2$, these values changes by at most 0.1 magnitudes. As in
previous plots, the size of the error bars represent cosmic variance
errors over the four fields.

We find that in the redshift range $0.4<z<1.2$ probed by our survey,
early-type galaxies are \textit{always} more clustered than late type
galaxies as we have already found in in the previous
Sections. Moreover, the difference in clustering amplitudes between
these two populations is approximately constant with redshift. (Note
that the galaxy samples examined here correspond to essentially the
brightest bins plotted in Figures~\ref{fig:mz_absmag_lz} and
Figure~\ref{fig:mz_absmag_hz}). We also find that the clustering of
the $-22 < M_B-5\log h < -19$ luminosity-limited full galaxy sample
(i.e., without including a type selection) decreases steadily from
$z\sim0.4$ to $z\sim1.2$.
\subsection{Relative bias}

We can also compute the relative bias between different galaxy
populations at different redshifts. At each redshift range in
Section~\ref{sec:lumin-deped-clust} ($0.2<z<0.6$ and $0.7<z<1.1$) we
compute the relative bias $b$ as follows:

\begin{equation}
b = b_a/b_b = \sigma_8(a)/\sigma_8(b)
\end{equation}

We adopt the usual definition for $\sigma_8$ \citep{P80}, 

\begin{equation}
\sigma_8=\sqrt{(C_\gamma(r_0/8)^{\gamma}), }
\label{eq:relbias}
\end{equation}

Where is $C\gamma$ is a constant which depends on $\gamma$:

\begin{equation}
C_\gamma={72\over(3-\gamma)(4-\gamma)(6-\gamma)2^\gamma}
\end{equation}

\begin{figure}
\resizebox{\hsize}{!}{\includegraphics{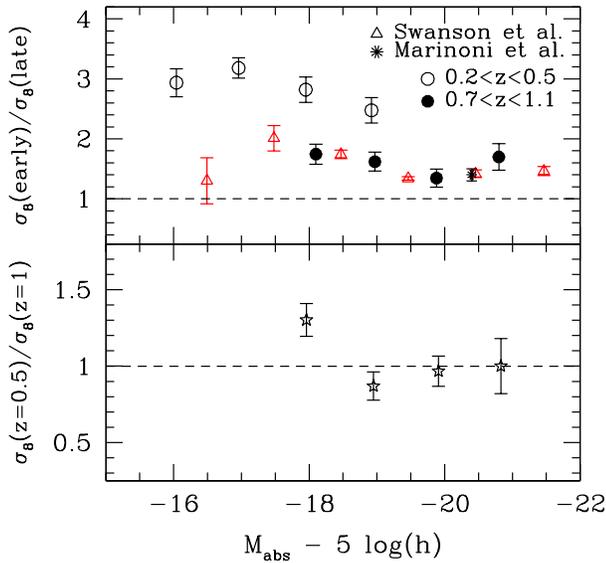}}
\caption{Top panel: relative bias between early and late-type
  populations at $z\sim0.5$ (open circles) and $z\sim1$ (filled
  circles). The open triangles and star show measurements from
  \cite{2007astro.ph..2584S} and \cite{2005A&A...442..801M}
  respectively. Bottom panel: relative bias between the low- and
  high-redshift full galaxy sample. For both panels, biases are
  plotted as a function of the median absolute rest-frame magnitude in
  slices of one magnitude in width.}
\label{fig:bias}
\end{figure}

In Figure~\ref{fig:bias} we plot the relative bias between the early
and late-type populations for our low and high redshift samples (open
and filled circles respectively) as a function of absolute rest-frame
luminosity. Error bars are computed from the field-to-field
variance. From our data it is clear that the relative bias between the
early and late type populations declines between $z\sim 0.5$ and
$z\sim1$.

We compare our relative bias between blue and red galaxy populations
with measurements in the literature. \cite{2005A&A...442..801M}
divided their spectroscopic sample into a red one with rest-frame
$(B-I)>0.95$ and a bluer one with $(B-I)<0.68$. By measuring the
probability distribution function (PDF) of these two populations, they
were able to be measure the relative bias of relatively bright ($M_B-5
\log (h)\sim -21$) galaxies in the interval $0.6<z<1.1$, indicated by
the starred point in Figure~\ref{fig:bias}. Recently,
\cite{2007astro.ph..2584S} investigated the relative bias between red
and blue spectroscopic galaxy samples in the Sloan Digital Sky Survey
as a function of absolute rest-frame red magnitude. Their results are
represented in Figure~\ref{fig:bias} as open triangles (with an
approximate offset applied to convert to rest-frame blue magnitudes
used in this paper).

From Figure~\ref{fig:bias} we see that our measurements at $z\sim1$
agree with \citeauthor{2007astro.ph..2584S} and
\citeauthor{2005A&A...442..801M}. \citeauthor{2007astro.ph..2584S}
findings are similar to ours: they observe an increase in the relative
bias between early and late types at fainter magnitudes. However, at
$z\sim0.5$, our relative bias measurements are considerably higher
than either measurements $z\sim0$ or $z\sim1$.

\section{Discussion}
\label{sec:disc-results-comp}

In this paper we have used a sample of 100,000 photometric redshifts
in the CFHTLS legacy survey deep fields to investigate the dependency
of galaxy clustering on rest-frame colour, luminosity and
redshift. The first sample we considered comprises a series of
magnitude-selected cuts sampling the full galaxy population from
$0.2<z<1.2$. We find that the clustering amplitude decreases gradually
from $3h^{-1}$~Mpc at $z\sim0.3$ declining to $2h^{-1}$ at
$z\sim1.0$. The declining correlation amplitude for the full galaxy
population at least to $z\sim1$ indicates that the field galaxy
population must be weakly biased, as this trend follows that of the
underlying correlation amplitudes of the dark matter.

We next repeated the same experiment (measuring galaxy clustering in
narrow redshift bins) and imposed an additional selection by absolute
luminosity and rest-frame colour. Selecting galaxies by slices of
absolute luminosity the steady decline in comoving correlation length
found for samples selected in apparent magnitude is even more
pronounced (Figure~\ref{fig:mz_absmag_new}). The luminous field galaxy
population, dominated by blue star-forming galaxies at $z\sim1$, is
clearly only weakly biased with respect to the dark matter
distribution.

Turning to rest-frame colour-selected samples at all redshift ranges
we consistently find that galaxies with redder rest-frame colours are
more strongly clustered than those with bluer rest-frame colours
(Figure~\ref{fig:mz_absmag_new}). Such an effect has long been
observed for galaxies in the local Universe \citep[for
example][]{2002MNRAS.332..827N,2005ApJ...630....1Z,1999MNRAS.310..281L,1997ApJ...489...37G}
and at higher redshifts for samples selected by type and luminosity
\citep{2006A&A...452..387M,2006ApJ...644..671C}. Numerical simulations
find a similar effect: For example, \citet{2004ApJ...601....1W} show
that older, redder galaxies are more strongly clustered. This is a
generic prediction from most semi-analytic models and hydrodynamic
simulations of galaxy formation: older, more massive galaxies formed
in regions which collapsed early in the history of the Universe. At
the present day such regions are biased with respect to the dark
matter distribution.

For the brightest ellipticals ($-22<M_B-5\log h<-19$) in our survey,
we find that their clustering amplitude does not change with redshift
(Figure~\ref{fig:mz_absmag_new}), indicating that at $z\sim1$ the
elliptical population must be strongly biased with respect to the
underlying dark matter distribution. Comparing our measurements for
objects with redder rest-frame colours with those of other surveys, we
find similar clustering amplitudes. Reassuringly, as we demonstrated
in Section~\ref{sec:clust-bright-early} sub-samples of galaxies with
redder rest-frame colours produce even higher correlation amplitudes
(Figure~\ref{fig:comparison}).

In a second set of selections we considered the dependence of
galaxy clustering on luminosity and type in two broad redshift bins:
$0.2<z<0.6$ and $0.7<z<1.1$ (we leave a 'gap' in the range $0.6<z<0.7$
to ensure that there is no contamination between high and low redshift
ranges). Once again, for the most luminous objects ($M_B-5\log h
\sim-20$) the correlation amplitude is approximately constant between
these two redshift bins. However, for fainter red objects, at a fixed
absolute luminosity, we see a decline in correlation amplitude between
$z\sim0.4$ and $z\sim1$; the same is true for samples selected purely
by absolute magnitude.  We find no evidence for a change in clustering
amplitude at the same luminosity for the blue population with
redshift.

At $0.2<z<0.6$, where we are complete to $M_B-5\log h~<-17$, we find
that red galaxies with $M_B-5\log h \sim -20$ are more strongly
clustered than bluer galaxies of the same luminosity. Moreover as the
sample rest frame luminosity decreases to $M_B-5\log h -18$ the
clustering amplitude rises from $\sim 4h^{-1}$~Mpc to $\sim
6h^{-1}$~Mpc. A similar effect has been reported in larger,
low-redshift samples in the local universe
\citep{2007astro.ph..2584S,2002MNRAS.332..827N}, where both the Sloan
and 2dF surveys have found higher clustering amplitudes for redder
objects fainter than $L^*$. Some evidence for this effect has also
been reported in numerical simulations \citep{2007MNRAS.374.1303C},
which indicate that this behaviour arises because faint red objects
exist primarily as satellite galaxies in halos of massive, strongly
clustered red galaxies. This means that less luminous, redder objects
reside primarily in higher density environments at $z\sim0.5$. This is
in agreement with recent studies of galaxy clusters at intermediate
redshift which indicate a rapid build-up of low luminosity red
galaxies in clusters since $z\sim1$ \cite{2007ApJ...670..206V}. Our
survey is not deep enough to probe to equivalent luminosities at
$z\sim1$.

Conversely for the redshift bin at $0.7<z<1.1$ we see that for the
full galaxy population more luminous objects are more strongly
clustered: $\sim 2 h ^{-1}$~Mpc for galaxies with $M_B-5\log h \sim
-19$ and $\sim 4 h ^{-1}$~Mpc for galaxies with $M_B-5\log h \sim
-21$. At all luminosity bins, galaxies with redder rest-frame colours
are always more strongly clustered than bluer galaxies.

In both redshift ranges, we measured the slopes of the correlation
function as a function of redshift, luminosity and rest-frame
colour. At $z\sim 1$ we observe that redder galaxies have steeper
slopes; at lower redshifts however different galaxy populations have
identical slopes. At these redshifts, we find steeper slopes in our
most luminous bin; at higher redshifts, no such obvious trend is
apparent (in contrast with \cite{2006A&A...451..409P}, who saw a clear
dependence of slope on absolute luminosity).

We have also computed the relative bias between red and blue galaxies
at $z\sim1$ and $z\sim0.5$. At $z\sim1$ our results agree with
measurements in the literature.  Our measurements at $z\sim0.5$ are
significantly above measurements made at $z\sim0$. Interestingly, our
results show that the relative bias between early and late types
increases gradually for samples selected with fainter intrinsic
luminosities, which is consistent with the results presented for our
investigation of galaxy clustering.

Concluding, we may summarise our results as follows: firstly, for
samples of galaxies with similar absolute luminosities, galaxies with
redder rest-frame colours are \textit{always} more strongly clustered
than their bluer counterparts. Secondly, for the bluer galaxy
populations, the correlation length depends only weakly on absolute
luminosity. At lower redshifts, we find some evidence that redder
galaxies with lower absolute luminosities are more strongly
clustered. For the \textit{entire} galaxy population (red and blue
types combined) we find that as the median absolute magnitude
increases, the overall clustering amplitude increases. For our the
most luminous red and blue objects, the clustering amplitude does not
change with redshift.

The overall picture we draw from these observations is that the
clustering properties of the blue population is remarkably invariant
with redshift and intrinsic luminosity. In general, galaxies with
bluer rest-frame colours, which comprise the majority of galaxies in
our survey, have lower clustering amplitudes (typically, $\sim 2
h^{-1}$~Mpc) than the redder populations. The clustering amplitude of
the blue population depends only weakly on redshift and
luminosity. This is consistent with a picture in which bluer galaxy
types exist primarily in lower density environments.

In contrast, the clustering amplitude of the low-luminosity red
population is lower at higher redshifts. In
Figure~\ref{fig:mz_absmag_new} we see that for the luminous
($M_B-5\log h \sim -20$) red population, the correlation amplitude
does not change with redshift. Moreover, at a fixed absolute
luminosity, the correlation amplitude of the full galaxy population
and the magnitude-selected galaxy population decreases from $z\sim
0.4$ to $z\sim1.1$, in step with the underlying dark matter
distribution.

\section{Conclusions}
\label{sec:conclusions}

We have presented an investigation of the clustering of the faint
($i'<24.5$) field galaxy population in the redshift range
$0.2<z<1.2$. Using 100,000 precise photometric redshifts extracted in
the four ultra-deep fields of the Canada-France Legacy Survey, we
construct a series of volume-limited galaxy samples and use these to
study in detail the dependence of the amplitude $A\_w$ and slope
$\delta$ of the galaxy correlation function $w$ on absolute $M_B$
rest-frame luminosity, redshift, and best-fitting spectral type (or,
equivalently, rest-frame colour). Our principal conclusions are as
follows:

1. The comoving correlation length for all galaxies with
$-19<M_B-5\log h<-22$ declines steadily from $z\sim0.3$ to $z\sim1$.

2. At all redshifts and luminosity ranges, galaxies with redder
rest-frame colours have clustering amplitudes between two and three
times higher than bluer ones.

3. For both the red and blue galaxy populations, the clustering
amplitude is invariant with redshift for bright galaxies with
$-22<M_B-5\log h<-19$.

4. At $z\sim0.5$, less luminous galaxies with $M_B-5\log h\sim-19$, we
find higher clustering amplitudes of $\sim 6h^{-1}$~Mpc.

5. The relative bias between populations of redder and bluer
rest-frame populations increases gradually towards fainter magnitudes.

The main implications of these results is that although the full
bright galaxy population traces the underlying dark matter
distribution quite well (and is therefore quite weakly biased),
redder, older galaxies have clustering lengths which are almost
invariant with redshift, and must therefore by $z\sim1$ be quite
strongly biased. In addition, at $z\sim0.5$ there is some evidence
that fainter red objects are more strongly clustered than $\sim L^*$
galaxies at these redshifts. This is consistent with a picture in
which fainter red objects exist primarily as satellite galaxies in
galaxy clusters.

It is tempting to interpret our results in terms of studies which show
that the number density of massive, luminous galaxies evolves little
from $z\sim1$ to the present day
\citep{2006A&A...453L..29C,2006A&A...455..879Z,CSH2}. In our survey,
the clustering amplitudes of bright ellipticals are already 'fixed in'
at $z\sim1$. Most of the changes in the clustering amplitude occur in
the fainter galaxy population. However, a full understanding of the
processes at work here will require mass-selected samples covering a
larger interval in redshift. Such samples will become possible in the
near future with the addition of near-infrared data to the CFHTLS
survey fields.

\section{Acknowledgements}
\label{sec:acknowledgement}
This work is based in part on data products produced at TERAPIX
located at the Institut d'Astrophysique de Paris. H.~J. McCracken
wishes to acknowledge the use of TERAPIX computing facilities. This
research has made use of the VizieR catalogue access tool provided by
the CDS, Strasbourg, France.


\end{document}